\documentclass{aa}

\usepackage{graphicx}
\usepackage{txfonts}
\usepackage{hyperref}
\usepackage{epstopdf}
\begin{document}

    \title{{3D simulations} of rising magnetic flux tubes in a compressible 
           rotating interior: The effect of magnetic tension}

    \titlerunning{Effect of magnetic tension on the rise of flux tubes}

    \author{Y. Fournier
       \and R. Arlt
       \and U. Ziegler
       \and K.G. Strassmeier}

    \institute{Leibniz-Institut f\"ur Astrophysik
     Potsdam (AIP), An der Sternwarte 16, D-14482 Potsdam, Germany}

    \date{Received \dots; accepted \dots}

\abstract
  %
  {Long-term variability in solar cycles represents a challenging
   constraint for theoretical models. 
   Mean-field Babcock-Leighton dynamos that
   consider non-instantaneous rising flux tubes have been shown to
   exhibit long-term variability in their magnetic cycle. 
   However a relation that parameterizes the {rise-time} of 
   non-axisymmetric magnetic flux tubes in terms of stellar parameters 
   is still missing.} 
  %
  {We aim to find a general parameterization of the {rise-time} of
   magnetic flux tubes for solar-like stars.}
  %
   {By considering the influence of magnetic tension on the rise of
    non-axisymmetric flux tubes, we predict the existence of a
    control parameter referred {as} $\Gamma_{\alpha_1}^{\alpha_2}$.
    This parameter is a measure of the balance between rotational 
    effects and magnetic effects (buoyancy and tension) acting on the 
    magnetic flux tube.
    We carry out two series of numerical experiments (one for
    axisymmetric rise and one for non-axisymmetric rise) and demonstrate
    that $\Gamma_{\alpha_1}^{\alpha_2}$ {indeed} controls the rise-time
    of magnetic flux tubes.}
  %
  {
   We {find} that the {rise-time} follows a power law of
   $\Gamma_{\alpha_1}^{\alpha_2}$ with an exponent that depends on 
   the azimuthal wavenumber of the magnetic flux loop.} 
  %
  {Compressibility does not impact the rise of magnetic flux tubes,
   while non-axisymmetry does. 
   In the case of non-axisymmetric rise, the tension force modifies
   the force balance acting on the magnetic flux tube. 
   We identified the three independent parameters required to predict
   the {rise-time} of magnetic flux tubes, {that is,}
   the stellar rotation rate, the magnetic flux density of the flux
   tube{, } and its azimuthal wavenumber. We combined {these}
   into one single relation that is valid 
   for any solar-like star. We suggest {using} this generalized
   relation to constrain the {rise-time} of magnetic flux tubes
   in Babcock-Leighton dynamo models.} 


\keywords{ magnetic fields -- sun: activity -- stars: magnetic field }

\maketitle

\section{Introduction}
\label{sec:introduction}

Solar and stellar dynamo theory is a complex field {that has seen} a lot of 
progress in recent decades \citep[{e.g.,} review by][]{Charbonneau14}.
{It has} also benefited  
from comparisons with disk-resolved stellar spot data \citep[e.g. review by][]{Strassmeier09},
but still lacks a universal solution for all stars possessing convective 
envelopes. In many stars, differential 
rotation is likely responsible for the conversion of poloidal into
toroidal fields. However the regeneration of poloidal magnetic field
is a more subtle issue \citep{Radler+03, Miesch05, Charbonneau10}. 

At least two types of dynamo {model} offer an explanation. 
The turbulent dynamo theory suggests that there is a net
effect from the turbulent electromotive force on large scales of which
one part -- the $\alpha$-effect -- provides the necessary regeneration  
\citep{Brun+04,Kapyla+10,Kapyla+12,Brown+10,Racine+11}. 
The Babcock-Leighton dynamo model (BL-dynamo) on the contrary,
describes a regeneration mechanism taking place near the surface where
magnetic flux emergence plays a major role \citep{Babcock62, Leighton69}. 
In many of those dynamo solutions, strong toroidal magnetic fields
reside near the tachocline. We suppose that these fields form magnetic
flux tubes. The concentration of magnetic flux is a special issue and
is {beyond} the scope of this paper. We focus on the properties of the
rise of such magnetic structures from the tachocline to the
surface. Hence, the present work addresses the concept of a BL-dynamo.  

BL-dynamos require active regions to regenerate the poloidal
field. The formation of active regions is also a debated issue
\citep{Cheung+14}. We recognize two main concepts here: {The}
magnetic flux tube model and the 
local formation model. The latter received an interesting
incentive through the Negative Effective Magnetic Pressure Instability
(NEMPI) which could be responsible for the formation of active regions
directly at the surface \citep[][and references therein]{Warnecke+13}. 
Within the same concept, we should also underline the local convective
model \citep{Rempel+14} which suggests the formation of active regions
due to local convective motion and granulation. In contrast to the
local formation, the magnetic flux tube models all require
coherent magnetic structures preceding the emergence. {The
present document discusses this scenario.}  

Some mechanisms have been suggested to form magnetic flux tubes. They
could be the manifestation of the concentration of magnetic flux due
to turbulence in the bulk of the convection zone
\citep{Nelson+14}. Alternatively, they could form at the tachocline
due to an entropic
instability in a magnetic layer sitting at the bottom of the
convection zone
\citep{Cattaneo+87,Cattaneo+89,Matthews+95,Hughes+97,Schussler+02}. We
suppose that flux tubes form as a result of the destabilization of a
magnetic layer in an unstable equilibrium against some kind of
Rayleigh-Taylor instability \citep{Wissink+00, Fan01}. 

In the present paper, we follow and discuss the idea of
\citet{Schussler80} and assume that stellar cycles as well as the
dynamo can be explained by rising flux tubes formed in the
tachocline. We study rising flux tubes in compressible rotating
stellar interiors with highly resolved, three-dimensional numerical
simulations. 
The issue of rising flux tubes was first addressed by \citet{Parker55}
who described the buoyant instability and the fact that flux tubes
could rise as coherent structures through the turbulent convection
zone. {This} idea became really attractive{, however,} after \citet{Spruit81}
derived the equations for the {thin-flux-tube} approximation. 
The approximation turned out to be a very useful description for
numerical experiments with a high level of precision.
This idea {became enriched by an increasing amount of}
physics and a complex model was presented recently by
\citet{Weber+15} which incorporates most of the relevant physics into 
the {thin-flux-tube approximation. 
This latter publication concluded} a long series of work which addressed various aspects of
the rise of magnetic flux tubes; the influence of initial conditions  
\citep{Spruit+82,Moreno-insertis83,Yoshimura85,Moreno-insertis+92,Fan+94},
and the need of rotation to reproduce the observations
\citep{Choudhuri+87,Van_ballegooijen83,Schussler+92,Caligari+94,Caligari+96,Deluca+97,Granzer+00,Granzer04}. 

In particular, \citet{Schussler+92} showed that in order to reproduce
large polar spots on short-period stars, the model {needed} to take
rotation into account. They also showed that several properties of the
rise of a flux tube scale with the magnetic Rossby number, 
%
\begin{equation}
  \label{rom-def}
  {\rm Ro}_{\rm m} = \frac{v_{\rm A}}{2 H_{\!P} \Omega} ~~\rm{,}
\end{equation}
%
where $v_{\rm{A}}=B/\sqrt{\mu_0\rho}$ is the Alfv\'en speed of a
magnetic field $B$ in a medium with density $\rho$, while $\mu_0$ is
the magnetic permeability, $H_{\!P}$ is the pressure scale height, and
$\Omega$ is the angular velocity. This is an important result if one
wants to learn from other stars, as it defines the regime of the
rise. 
One of the aims of the present work is to discuss this result and {to} try
to generalize it to the non-axisymmetric case. 

Finally, it has been kept in mind that the {thin-flux-tube}
approximation has limits \citep{Cheung+06}. This is why
Boussinesq as well as anelastic rising flux tubes (``thick flux
tubes'') were simulated in axisymmetry {(two-dimensional (2D))}
\citep{Moreno-insertis83,Moreno-insertis+97,Choudhuri+87,Chou+89,Fan+98}.  
Getting rid of the thin aspect of the flux tube {introduced the need for}
twist to maintain the coherence of the magnetic structures along their
rise 
\citep{Browning+83, Moreno-insertis+96, Longcope+97, Moreno-insertis+97}.

Later, \citet{Fan08} showed that the azimuthal asymmetry of the rise of
a magnetic flux tube influences its dynamics. A variety of papers {have}
been published on non-axisymmetric {(three-dimensional (3D))} studies 
\citep{Jouve+09, Weber+11, Fan+13, Pinto+13, Weber+15}. 

We present the first fully compressible non-axisymmetric 
model of rising flux tubes in a rotating stellar interior. We
conducted parameter studies which enable us to extend the control parameter
(\ref{rom-def}) for the axisymmetric case to the {3D} case.
The original idea consists in challenging
the flux tube theory, and incorporating our results into a mean-field
dynamo model{, } which could be validated by observations and could
help to constrain the flux tube dynamo theory. In 
{Section}~\ref{sec:equation-and-setup} we first describe the numerical
experiments. Then
{in Section}~\ref{sec:scaling-relation-of-rising-flux-tube} we derive the
scaling parameter from first principles and make a prediction for the
non-axisymmetric rise. Before studying the non-axisymmetric case
we first validate the 2D setup 
{in Section}~\ref{sec:validation-of-the-setup-with-a-2d-numerical-experiment}
by comparing our results with the literature. In 
{Section}~\ref{sec:extension-to-the-non-axisymmetric-case-3d}, we show
that our theoretical predictions for non-axisymmetric rises are
valid. Finally, we discuss the limits of our results and possible
improvements {in Section}~\ref{sec:discussions}. In the Appendix, we
enclose the derivation of the stratified interior and the full list of
our simulations.

\section{Equations and setup}
\label{sec:equation-and-setup}

Our study relies on 2D and 3D numerical simulations. Since most of the
presented simulations differ by just one parameter, we often refer to
some standard cases. For each simulation we specify the modified
parameter presupposing that all others are taken from the standard
model. 
From the end of this Section we refer to a fiducial 2D simulation
(STD-2D), and a standard 3D simulation (STD-3D).

\subsection{Compressible MHD set of equations}
\label{sec:1--compressible-mhd-set-of-equations}

We are interested in rising magnetic flux tubes in stellar
interiors. While in the present paper, we focus on solar-like
stars, our model is constructed such that we could cover more than
this specific case and extend our study to other dwarfs and red 
giants in the future. We solve the fully compressible, resistive MHD equations. One 
could argue that in solar-like stars the anelastic approximation is
sufficient and saves a lot of computational time. But in case of red
giants this assumption may not hold anymore.

The MHD equations as we solve them can be written in the following
compact form:  
%
\begin{equation}
  \partial_t \rho = -\nabla\cdot(\rho \vec{u}) ~~\rm{,}
\end{equation}
\begin{eqnarray}
  \partial_t (\rho \vec{u}) &=&- \nabla \cdot \left[ \rho \vec{u} \vec{u} 
                                +  P_{\rm{tot}} I 
                                -  \frac{1}{\mu_0} \vec{B}\vec{B} \right] 
                                   \nonumber\\ 
                            & &+ \rho \vec{g} + \rho \vec{f} ~~\rm{,} 
                                   \nonumber\\ 
\end{eqnarray}
\begin{eqnarray}
  \partial_t e &=&- \nabla \cdot \left[ (e +P_{\rm{tot}}) \vec{u} 
                   -  \frac{1}{\mu_0}(\vec{u} \cdot \vec{B})\vec{B} \right] 
                      \nonumber\\
               & &+ \nabla \cdot \left[\frac{\eta}{\mu_0}\vec{B} 
                      \times (\nabla \times \vec{B} ) 
                    - \vec{F}_{\rm{rad}} \right] 
                      \nonumber\\ 
               & &+ \rho \vec{g} \cdot \vec{u} + \rho \vec{f} \cdot \vec{u} 
                      ~~\rm{,} \nonumber\\ 
\end{eqnarray}
\begin{equation}
  \partial_t \vec{B} =  \nabla \times ( \vec{u} \times \vec{B} ) 
                        + \eta\nabla^{2} \vec{B} ~\rm{,}
\end{equation}
\begin{equation}
  P = \frac{\rho k_{\rm B}T}{m\mu} ~~\rm{.}
\end{equation}
%
Here, $\vec{u}$, $P_{\rm{tot}}$, $\vec{B}$, $\vec{g}$, $\vec{f}$, $T$,  
$\vec{F}_{\rm{rad}}$, $k_{\rm B}$, $m$ and $\mu$ are the velocity
field, the total pressure being the sum of the thermal ($P$) and the
magnetic pressure ($P_{\rm m}=B^2/2\mu_0$), magnetic flux density,
gravitational acceleration, external acceleration (namely Coriolis),
the temperature, the radiative flux, the Boltzmann constant, the
atomic mass unit and the mean molecular weight, respectively. Apart
from the usual symbols, $\partial_{t}(.)$ and $I$ are the partial time
derivative and the identity matrix, respectively.  

\begin{figure}[bpth!]
  \centering
  \includegraphics[width=1.0\linewidth]{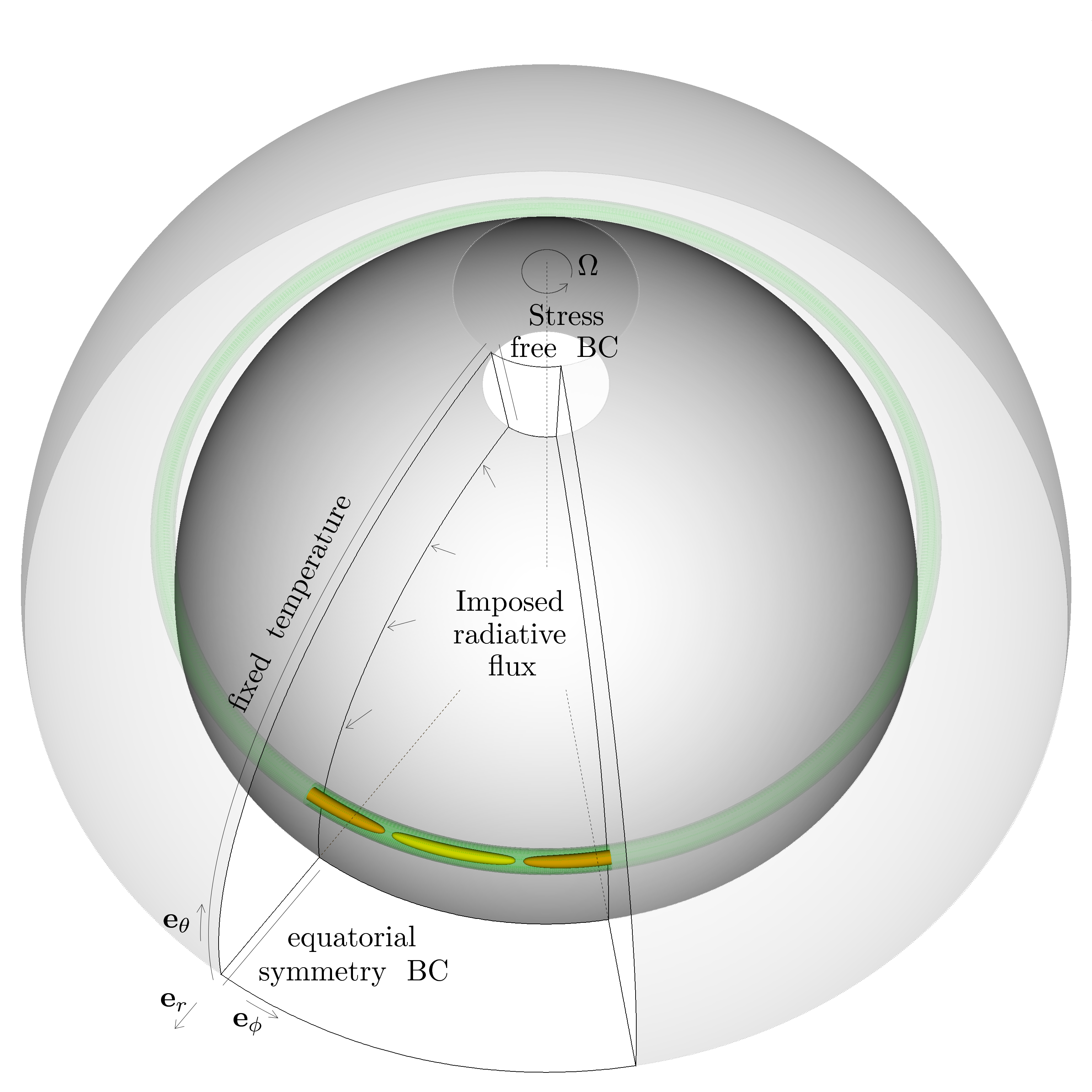}
  \caption{\label{fig:full-dom} Sketch of the numerical setup. The
           yellow contour shows the buoyant part of the flux tube
           with a lack of entropy. This section of the flux tube will
           buoyantly rise toward the surface and emerge as an active
           region. The orange contour surfaces show the
           neutrally buoyant
           parts of the flux tube, with an excess of entropy. The
           latter sections will remain rooted at the bottom of the
           simulated domain shown by the wedge-like shape of solid
           lines. The azimuthal morphology shown here is valid only in
           3D, for an azimuthal wavenumber of $m=8$. In 2D, the flux
           tube is buoyant everywhere,
           and the computational domain is a meridional plane. The
           direction of rotation is also indicated.} 
\end{figure}

We solve the equations in a fraction of a spherical shell with varying
azimuthal extent depending on the needs of various non-axisymmetries
(Fig.~\ref{fig:full-dom}). This system is solved on a spherical grid
with the parallelized NIRVANA code, described in detail by
\citet{Ziegler11} 
\footnote{http://www.aip.de/Members/uziegler/nirvana-code}. The
spherical coordinate system is $(r, \theta, \phi)$. 

\subsection{Rotating adiabatically-stratified stellar interior}
\label{sec:1--rotating-adiabatically-stratified-stellar-interior}

We study the rise of magnetic flux tubes in a hydrostatic,
adiabatically stratified layer. 
The design and the analysis of a convective zone is by itself a
complex problem and should be addressed separately. We choose to first
tackle the issue of a non-convective environment. In this situation,
the radiative flux transports the entire luminosity radially along the
unit vector $\vec{e}_r$:  
%
\begin{equation}
  \vec{F}_{\rm rad} = - \kappa \frac{dT}{dr} \vec{e}_r,
\end{equation}
%
where $\kappa$ is the thermal conductivity. 
This layer is also hydrostatic, the gradient of pressure balances
gravity (here approximated by a point mass), 
%
\begin{equation}
  \frac{dP}{dr} = - \rho \frac{GM_\star}{r^2},
\end{equation}
%
where $M_\star$ is the stellar mass. We defined the pressure scale
height at the top of the domain as {follows}:
%
\begin{equation}
  H_{\!P_0} = P_0 ~ \left( \left. \frac{dP}{dr} \right|_{R_0} \right)^{-1},
\end{equation}
%
where $P_{0}$ and $R_{0}$ are the pressure and the radius at the top
of the domain, respectively. The logarithmic temperature gradient is 
%
\begin{equation}
  \nabla = \frac{d \ln T}{d \ln P} ,
\end{equation}
%
where the special case of $\nabla=0.4$, referring to an adiabatic
stratification, is denoted by $\nabla_{\rm ad}$. Making use of these
four equations one can obtain the analytical expression for the three
thermodynamic quantities $T$, $\rho${, } and $P$. 
%
\begin{figure}[bpth!]
  \centering
  \includegraphics[width=1.0\linewidth]{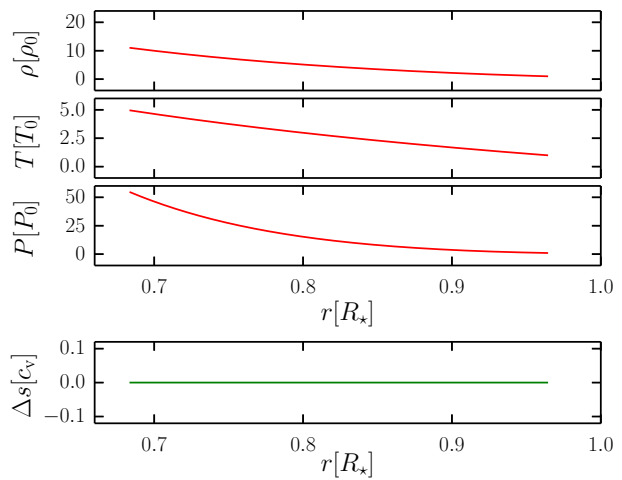}
  \caption{\label{fig:atmosphere}Radial profiles of $\rho$, $T${, } and $P$, 
           normalized by their respective values at the top of the
           domain. The radial profile of the normalized gradient of
           entropy is a constant (adiabatic) in this configuration.}
\end{figure}
%
\begin{eqnarray}
  T(r) & = & T_0 \left[ 1 + \frac{\nabla R_0^2}{H_{\!P_0}} 
                            \left(\frac{1}{r} - \frac{1}{R_0}\right) 
                 \right]                                                ,\\
  P(r) & = & P_0 \left(\frac{T(r)}{T_0}\right)^{1/\nabla}               ,\\
  \rho(r) & = & \rho_0 \left(\frac{T(r)}{T_0}\right)^{1/\nabla-1}.
\end{eqnarray}
%
The index 0 refers to the quantities at the top boundary of the
domain. We illustrate
their profiles in Fig.~\ref{fig:atmosphere} where we also show the
constant entropy gradient profile characterizing adiabatic layers. 
For more details, please refer to the
Appendix. Furthermore, the entropy gradient is defined as 
%
\begin{equation}
  \Delta{s}(r) = c_{\rm{v}} \log \left( \frac{P(r)}{P_{0}} \right) 
               - c_{\rm{p}}  \log \left( \frac{\rho(r)}{\rho_{0}} \right),
\end{equation}
%
where $c_{\rm{v}}$ and $c_{\rm{p}}$ are the volumetric heat capacity
and the pressure heat capacity, respectively. 

A proper treatment of the top boundary condition requires at least a
few tens of grid points to resolve the pressure scale height at the
top of the domain ($0.964~R_\star$).
In case the pressure scale height is not sufficiently resolved, the
hydrostatic equilibrium cannot be maintained. 
Therefore stratification imposes a minimum radial resolution necessary
for the simulation.

\subsection{Boundary conditions}
\label{sec:1--boundary-conditions}

Both models, STD-2D and STD-3D, have similar boundary conditions. As
illustrated in Fig.~\ref{fig:full-dom}, boundary conditions for the
{thermodynamical} variables are a constant thermal flux at the bottom of
the convection zone, and a constant temperature $T_0$ at the top of
the domain. In the latitudinal direction, we choose zero thermal flux 
condition at the equator and at high latitude. 
For the momentum equation, all boundaries are stress-free. 
For the STD-3D setup we apply periodic boundary conditions in the 
azimuthal direction. For the magnetic field, we use a pseudo-axis 
boundary condition for the
high-latitude boundary, a reverse boundary at the equator, and a
pseudo-vacuum at the inner and outer radial boundaries, {that is,} a
vertical-field condition.

\subsection{Useful parameters}
\label{sec:equations-and-setup--useful-parameters}

{Throughout} this study we make use of several parameters. The
plasma-$\beta$ is the ratio of gas pressure to magnetic pressure,
%
\begin{equation}
  \beta = P/P_{\rm m} ~{\rm .}
\end{equation}
%
The rotational Mach number is used to describe rotation, defined as
the ratio of the rotation speed to the sound speed $c_{\rm{s}}$, 
%
\begin{equation}
  {\cal M}_{\rm{rot}} = \frac{\varpi_{\rm ini} \Omega}{c_{\rm{s}}} ~{\rm ,}
\end{equation}
%
where $\varpi_{\rm ini}$ is the axis distance of the initial tube
location, $(r_{\rm ini},\theta_{\rm ini})$, {that is, } $\varpi_{\rm ini}=r_{\rm ini}\sin\theta_{\rm ini}$. 
We also introduce the Lorenz number (${\rm Lo}$) according to 
%
\begin{equation}
  {\rm Lo} ~=~ \frac{ v_{\rm A} }{\varpi_{\rm ini}\Omega} 
           ~=~ \frac{2 H_{\!P} }{\varpi_{\rm ini}} 
             ~ {\rm Ro}_{\rm m} ~{\rm .}
\end{equation}

We further introduce the relative {rise-time} defined as
%
\begin{equation}
  \tilde{\tau}_{\rm{rise}} = \frac{\tau_{\rm{rise}}}{P_{\rm{rot}}} ~{\rm ,}
\end{equation}
%
where $\tau_{\rm rise}$ and $P_{\rm rot}$ are the time needed by the
flux tube to reach the emergence line ($0.95~R_{\star}$) of the
computational domain, and the rotation period, respectively. 

Numerical experiments can take advantage of dimensionless
variables. We present the dimensionless system used in our setup in
Table~\ref{tab:dimensionless}.
%
\begin{table}
  \caption{\label{tab:dimensionless} Dimensionless definition of main
           quantities. $G$ is the gravitational constant and $M_\star$
           the mass of the star. $t_{\rm ff}$ is the
           free-fall time-scale. 
          }
  \centering
  \begin{tabular}{rc}
    \hline\hline
    Quantity & Unit\\
    \hline
    & \\
    Density & $\rho_{0}$\\
    & \\
    Length & $R_\star$\\
    & \\
    Time & $\sqrt{\frac{R_\star^{3}}{GM_\star}} = t_{\rm{ff}}$\\
    & \\
    Pressure & $\rho_{0}\frac{GM_\star}{R_\star}$\\
    & \\
    Temperature & $\frac{GM_\star}{R_\star^{3}}$\\
    & \\
    Entropy & $c_{\rm{v}}$\\
    & \\
    Magnetic field & $B_{\rm eq} = \sqrt{ 2 \rho_{0} \mu_{0} \frac{GM_\star}{R_\star}}$\\
    & \\
    \hline
  \end{tabular}
\end{table}
%
The dimensionless pressure scale height at the top of the domain is
defined by  
%
\begin{equation}
  \frac{H_{\!P_{0}}}{R_{\star}} = \chi ~{\rm .}
\end{equation}

\subsection{Criterion for adaptive mesh refinement (AMR)}
\label{sec:1--criterium-for-adaptive-mesh-refinement}

Active-region latitudes span from 0$^\circ$ to 40$^\circ$. Catching 
such a wide latitudinal band requires a considerable fraction of the
spherical domain. Therefore we need to investigate rising magnetic
flux tubes in global simulations.  
However, flux tubes are small coherent structures in this large
domain. 
Because of the strong stratification of the solar interior and
supposing that magnetic flux tubes conserve their internal flux all
along their rise, a 60~Mm active region can only result from the
emergence of a magnetic flux tube that had a radius less than 1\% of
the convective zone's radial {extent}, when it formed at the bottom of
the convective zone.
In addition, 
the numerical conservation of the magnetic flux in the tube {forces} us to
resolve the magnetic flux tube with at least 50~points in diameter. 

Adaptive mesh refinement (AMR) allows us to locally add resolution, 
and therefore meets the requirements of our situation by resolving
small structures in large domains \citep{Ziegler12}. The AMR procedure checks{, } at each
time step{, } the refinement (derefinement) condition: if a cell fulfills a 
given criterion, a resolution level is automatically added (removed). 
In 2D, a cell will be divided into four smaller cells{; } in 3D into
eight cells.  
We chose the presence of a magnetic field as a refinement condition:
if the strength of the magnetic field exceeds 10\% of the maximum
value in the domain, the grid will be refined. We limit the code to
two refinement levels on top of the base level to avoid over-refinement
and save {computation} time.

\subsection{Flux tube definition}
\label{sec:1--flux-tube-definition}

At the bottom of the prescribed static layer, we introduce a flux tube
in non-equilibrium. The flux tube is a twisted torus of constant
magnetic field along the azimuthal direction. The magnetic field
strength decreases with the distance from the tube's center at 
$(r_{\rm ini}, \theta_{\rm ini})$. The strength of the azimuthal
magnetic field is assumed to be
%
\begin{equation}
  B_{\phi}(r_{\rm ft}) = B_0 ~ \exp \left(-\frac{r^{2}_{\rm ft}}{ R^{2}_{\rm ft}} \right),
\end{equation}
%
where $r_{\rm ft}$, $R_{\rm ft}${, } and $B_0$ are the distance from
$(r_{\rm ini}, \theta_{\rm ini})$, the initial radius of the flux
tube, and the maximum strength of the magnetic flux tube,
respectively. The radius of the flux tube ($R_{\rm ft}$) defines the
torus{, } which contains about 98\% of the magnetic flux initially. $B_0$
is the magnetic field strength in the middle of the flux tube, 
and corresponds to the value used in {thin-flux-tube} approximation
models.  

\begin{figure}[bpth!]
  \centering
  \includegraphics[width=1.0\linewidth]{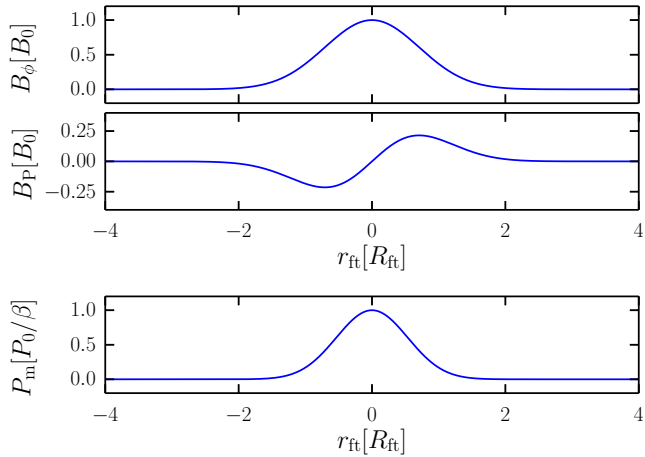}
  \caption{\label{fig:fluxtube} Profile of the dimensionless magnetic
           quantities over the normalized distance from the flux tube
           center. $B_{\phi}$, $B_{\rm{p}}${, } and $P_{\rm{m}}$ are the
           toroidal component of the twisted magnetic field, the
           poloidal component of the twisted magnetic field, and the
           resulting magnetic pressure{, } respectively.
          } 
\end{figure}

While a purely azimuthal magnetic field is by construction divergence
free, ensuring the solenoidality of a twisted flux tube is less
trivial. We therefore derive the field twist from 
a vector potential. This component is circular around the tube center,
with strength
%
\begin{equation}
  \frac{dA_{\phi}}{dr_{\rm ft}} = B_{\rm{p}}(r_{\rm ft}) 
                                = \lambda \frac{r_{\rm ft}}{R_{\rm{ft}}} 
                                  B_{\phi}(r_{\rm ft}) ~{\rm ,}
\end{equation}
%
where $\lambda$ is the twist parameter and $A_{\phi}$ is the azimuthal
component of the magnetic vector potential. Figure~\ref{fig:fluxtube}
shows the resulting radial profiles of $B_{\rm p}$ and $B_{\phi}$.
In the present paper, we fixed all characteristics of the flux tube
except $B_{0}$. Its value is computed from the input parameters.

We assume that flux tubes form at the tachocline. Therefore, we
set the initial depth $r_{\rm ini} = 0.71~R_\star$. Furthermore, 
we chose $\theta_{\rm ini} = 20\degr$ because the {thin-flux-tube}
theory demonstrated that such simulated axisymmetric flux tubes 
emerge at the observed active latitudes.

We set the initial size of the flux tube such that it meets two
requirements: {The} magnetic flux should correspond to the flux observed
in large emergence regions and the diameter of the flux tube should 
contain more than 50~grid~points.
Such a resolution is sufficient to ensure the numerical diffusion to
be negligible along the flux tube's rise; {and} the flux tube will conserve
about 90\% of its initial flux. 
The resolution we have chosen represents a
compromise between realistic physics and realistic computation time (see Table~\ref{tab:std-2d}).

In the simulations presented here, magnetic flux tubes rise
sufficiently {slowly} to consider them in pressure equilibrium;
the presence of magnetic pressure lowers the thermal
pressure inside the radius of the flux tube compared to the thermal
pressure of its surroundings. 

The thermal state of the
flux tube controls how the pressure lack determines the other
thermodynamical quantities, $\rho$ and $T$. The initial thermal state
of a flux tube has been shown to have great impact on the dynamics of
the rise \citep{Moreno-insertis83}. We illustrate two extreme
situations here: (a) an isothermal flux tube, and (b) a neutrally
buoyant flux tube.  

\begin{figure}[bpth!]
  \centering
  \includegraphics[width=1.0\linewidth]{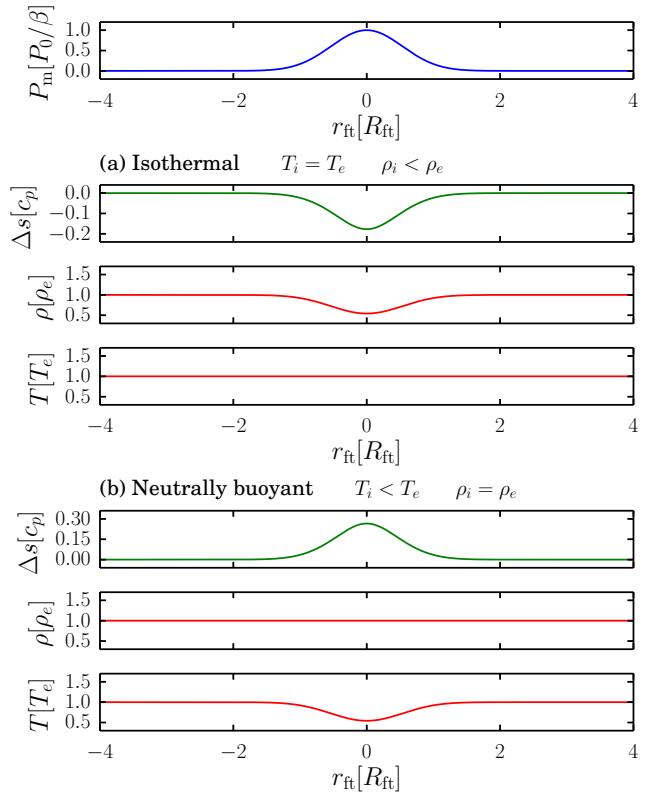}
  \caption{\label{fig:isothermal} Profiles of the magnetic pressure
           and the thermodynamical quantities for a magnetic flux tube
           in two extreme hydrostatic states, (a) in an isothermal
           state and (b) in a neutrally buoyant state.}
\end{figure}

In order to discuss these two different thermal states and for more
convenience, from now on, we use the indices $()_{\rm i}$ and 
$()_{\rm e}$ for quantities inside the flux tube at $r_{\rm ft} = 0$
and outside the flux tube at $r_{\rm ft} > 2 R_{\rm ft}$, respectively. 
As shown in Fig.~\ref{fig:isothermal}, the isothermal case 
is accompanied 
by a lack of density. In such a case, the flux tube
is buoyant. The neutrally buoyant case,
however, consists of a cool flux tube. The lack of thermal pressure
exclusively applies to temperature, allowing the densities inside and
outside the flux tube to be equal. We note that even if such
a flux tube is buoyantly neutral, it will still rise due to conduction
of heat inside the tube. However such a flux tube rises on a much
longer time scale. 

\subsubsection{The thermal state of the axisymmetric case: STD-2D}
\label{sec:1--the-thermal-state-of-the-axisymmetric-case}

In the axisymmetric case (azimuthal wavenumber $m=0$) the 
flux tube remains a torus {during the totality of} its ascent. The flux tube is
assumed initially isothermal{; } hence, it is
buoyant
everywhere. We denote the internal density by 
$\rho_{\rm i}^{(0)}$, where $()^{(m)}$ indicates the wavenumber of the
most unstable mode. The internal density can be simply written as
%
\begin{equation}
  \rho_{\rm i}^{(0)} = \rho_{\rm iso} \equiv \rho_{\rm e} 
                       \left( 1 - \frac{1}{\beta} \right).
  \label{rho_2d}
\end{equation}
%
{We note} that the internal density varies with the location of the flux
tube, since $\rho_{\rm e}$ comes from the stellar model and is a
function of radius $r$.  

\begin{table}
  \caption{\label{tab:std-2d} Fixed parameters for the numerical setup
           and the initial conditions of the STD-2D case. The last
           three parameters concern only the magnetic flux tube.} 
  \centering
  \begin{tabular}{rc}
    \hline\hline
    \textbf{Numerical parameters} & \\
    \hline
     & \\
         Resolution & $[ 512 \times 1024 ]$\\
         AMR levels & 2 levels\\
         Effective resolution & $[ 2048 \times 4096 ]$\\
     & \\
    \textbf{Domain definitions} & \\
    \hline
     & \\
         Radial domain ($R_\star$) & [ 0.684 -- 0.964 ]\\
         Latitudinal domain ($\degr$) & [ 0 -- 81 ]\\
         $\chi$                       & 0.013\\
         Stratification $\rho_{\rm bot}/\rho_{\rm top}$ & $\approx 50$\\
     & \\
    \textbf{Initial conditions} & \\
    \hline
     & \\
         Radius $r_{\rm ini}$ ($R_\star$)             & 0.71\\
         Latitude $\theta_{\rm ini}$                  & $20\degr$\\
         Initial radius $R_{\rm ft}$ ($R_\star$)      & $10^{-3}$\\
     & \\
    \hline
  \end{tabular}
\end{table}

\subsubsection{The thermal state of the non-axisymmetric case: STD-3D}
\label{sec:1--the-thermal-state-of-the-non-axisymetric-case}

In the non-axisymmetric case, the flux tube is initially an
axisymmetric torus. While rising{, } it evolves into an asymmetric
$\Omega$-shaped loop exhibiting some writhe. Since we 
envisage that flux tubes rise in {the} form of loops, we introduce a
periodic perturbation along the azimuthal coordinate
with an amplitude of $\rho_{\rm e} - \rho_{\rm iso}$,
so that an aziuthal section of the tube is buoyant
  (see the yellow contour in Fig.~\ref{fig:full-dom}).
The apex of the
loop rises, while the feet of the loop remain almost neutrally buoyant and
stay around $r_{\rm ini}$.
It is important to note that we do not impose any artificial anchoring
  on the feet of the loop. The resulting rising $\Omega$-shaped loop
  may writhe and tilt.

In contrast to the axisymmetric flux tube, we break the symmetry by an
entropic wave similar to the one described by \citet{Jouve+09}. 
The dependence of the internal density on the
azimuthal direction is denoted by $\rho_{\rm i}^{(m)}$ and is defined
by  
%
\begin{equation}
  \rho_{\rm i}^{(m)} = \rho_{\rm iso} 
                     + \frac{1}{2}\left(\rho_{\rm e} 
                     - \rho_{\rm iso} \right)
                       [1 - \cos(m \phi)].
\end{equation}
%
For $m=0$, the definition is equivalent to the STD-2D thermal state. 

We are simulating a spherical wedge with an azimuthal extent of
$\pi/4$ which is one-eighth of the full azimuthal range. In order to
obtain a single apex in the computed domain, we therefore choose to
give a wave number $m = 8$ to our entropic wave. 

The resolution, stratification, and flux tube radius are set to
account for the same constraints as in the 2D case. However since 3D
simulations are much more time consuming (320~times longer at the same
resolution), we had to reduce the resolution by four, leading to lower
stratification ($\approx 11$) and a ten times larger flux tube
radius. We think that even at this large increase in radius and at this
large decrease in resolution, the drag force is kept sufficiently
low so that we are still in a kinematic regime, the viscous force does
not dominate. Furthermore, AMR enables us to keep reasonable computing
times, retaining again about 90\% of the initial magnetic flux. 

We summarize the standard parameter values of the STD-3D model in
Table~\ref{tab:std-3d}.

\begin{table}
  \caption{\label{tab:std-3d} Fixed parameters for the numerical setup
           and the initial conditions of the STD-3D case. Again, the
           last three parameters concern only the magnetic flux tube.} 
  \centering
  \begin{tabular}{rc}
    \hline\hline
    \textbf{Numerical parameters} & \\
    \hline
     & \\
         Resolution & $[ 128 \times 256 \times 80 ]$\\
         AMR levels & 2 levels\\
         Effective resolution & $[ 512 \times 1024 \times 320 ]$\\
     & \\
    \textbf{Domain definitions} & \\
    \hline
     & \\
         Radial domain ($R_\star$) & [ 0.684 -- 0.964 ]\\
         Latitudinal domain ($\degr$) & [ 0 -- 81 ]\\
         Azimuthal domain ($\degr$) & [ 0 -- 45 ]\\
         $\chi$ ($R_\star$) & 0.04\\
         Stratification $\rho_{\rm bot}/\rho_{\rm top}$ & $\approx 11$\\
     & \\
    \textbf{Initial conditions} & \\
    \hline
     & \\
         Radius $r_{\rm ini}$ ($R_\star$)        & 0.73\\
         Latitude $\theta_{\rm ini}$             & $20\degr$\\
         Initial radius $R_{\rm ft}$ ($R_\star$) & $10^{-2}$\\
     & \\
    \hline
  \end{tabular}
\end{table}

\section{The effect of local magnetic tension in non-axisymmetric rise}
\label{sec:scaling-relation-of-rising-flux-tube}

We aim to find a parameter that controls the {rise-time} of magnetic
flux tubes in rotating stellar interiors from compressible numerical  
experiments. In contrast to anelastic or {thin-flux-tube} simulations, 
compressible experiments suffer from an upper limit for the $\beta$
parameter. Compressible codes are not making use of a background
state to solve the MHD equations, hence $\Delta\rho/\rho$ has to be
larger than the discretization error, whereas anelastic simulations
ensure this by construction. For the setup's resolution,
$\beta$ is limited to $\beta_{\rm max} \approx 200$ for STD-2D and
$\beta_{\rm max} \approx 50$ for STD-3D. In the solar case, however,  
$\beta$ is expected to be of the order of $10^{5}$. Hence, it is
crucial to be able to define the regime of the buoyant rise in order
to scale the results to higher $\beta$ and compare our results with
other simulations and observations. 

\cite{Choudhuri+87} underlined that the ratio of the buoyant force to
the Coriolis effect controls the regime of an axi\-symmetric buoyant rise,
%
\begin{equation}
  \label{eq:fbuoy-o-fcorio}
  \frac{F_{\rm buoy}}{F_{\rm corio}} ~=~ \frac{\Delta \rho ~ g}
                                     {2\rho ~ v_{\rm rise} ~ \Omega},
\end{equation}
%
where $\Delta \rho$ and $v_{\rm rise}$ are the lack of density inside
the flux tube and the rise velocity of the tube, respectively. When
this ratio exceeds unity, the regime of the rise is 
buoyancy dominated; when the ratio becomes less than unity the 
regime is rotation dominated. 

As \cite{Schussler+92} pointed out for axi\-symmetric simulations, 
the rise velocity, $v_{\rm rise}$, corresponds to the buoyant velocity: 
%
\begin{equation}
  v_{\rm buoy} ~=~ \left[ 2 \frac{ \Delta \rho}{\rho} 
                            g l \right]^{1/2} 
\end{equation}
%
with $\Delta \rho / \rho = 1/\beta$ according to Eq.~(\ref{rho_2d}) and
$l = H_{\!P}$ {being} the local pressure scale height. Replacing 
$v_{\rm rise}$ by $v_{\rm buoy}$ in Eq.~(\ref{eq:fbuoy-o-fcorio}), we
can rewrite the ratio as
%
\begin{equation}
  \frac{F_{\rm buoy}}{F_{\rm corio}} ~=~ 
      \left( \frac{3 \sqrt{\gamma}}{2} 
      \frac{ v_{\rm ff} }{ c_{\rm s}     } \right)
      \frac{ v_{\rm A}  }{ \varpi \Omega }  
     ~~\propto~ {\rm Lo},
\label{eq:fbuoy-o-fcorio-v}
\end{equation}
%
where $v_{\rm ff} = \sqrt{H_{\!P} g}$ is the free fall velocity and 
${\cal M}_{\rm ff} = v_{\rm ff} / c_{\rm s}$ is the free fall Mach
number. 
In the present series the latter {remains} constant and is
{approximately} unity. The
variables in Eq.~(\ref{eq:fbuoy-o-fcorio-v}) are $v_{\rm A}$ and $\varpi
\Omega$. Therefore, the ratio is proportional to the Lorentz
number. Two simulations with different parameters but the same
Lorentz number will deliver the same solution; this conclusion being
valid exclusively for axisymmetric simulations. 

  In a non-axisymmetric rise, magnetic tension will alter the ratio of
  forces. Due to the high $\beta$ in stellar interiors, magnetic tension never
  dominates, but because it is directed inwards it acts against
  the buoyant rise. The magnetic tension reduces the rise velocity
  which alters the Coriolis effect.

  Magnetic tension can be approximated by
%
\begin{equation}
  F_{\rm tens} ~=~ \frac{2 P_{\rm m} }{ {\cal R} } ,
\end{equation}
%
where ${\cal R}$ is the curvature radius of the rising flux tube.

In the axisymmetric case the curvature radius, $S{\cal R}$, is the distance of the flux 
tube from the rotation axis ($\varpi$) and it is sufficiently large {to make}
the tension force negligible. The regime depends on two
independent variables, $\beta$ and ${\cal M}_{\rm rot}$, which control the
buoyant force and the Coriolis effect, respectively.
In the non-axisymmetric case, flux tubes rise in the form of $\Omega$-loops. 
The curvature radius, ${\cal R}$, of the $\Omega$-loops, naturally connects
to the azimuthal wavenumber, $m$. Non-axisymmetry introduces a new degree of
freedom, that requires an additional parameter: $m$.

\begin{figure}[t]
  \centering
  \includegraphics[width=1.0\linewidth]{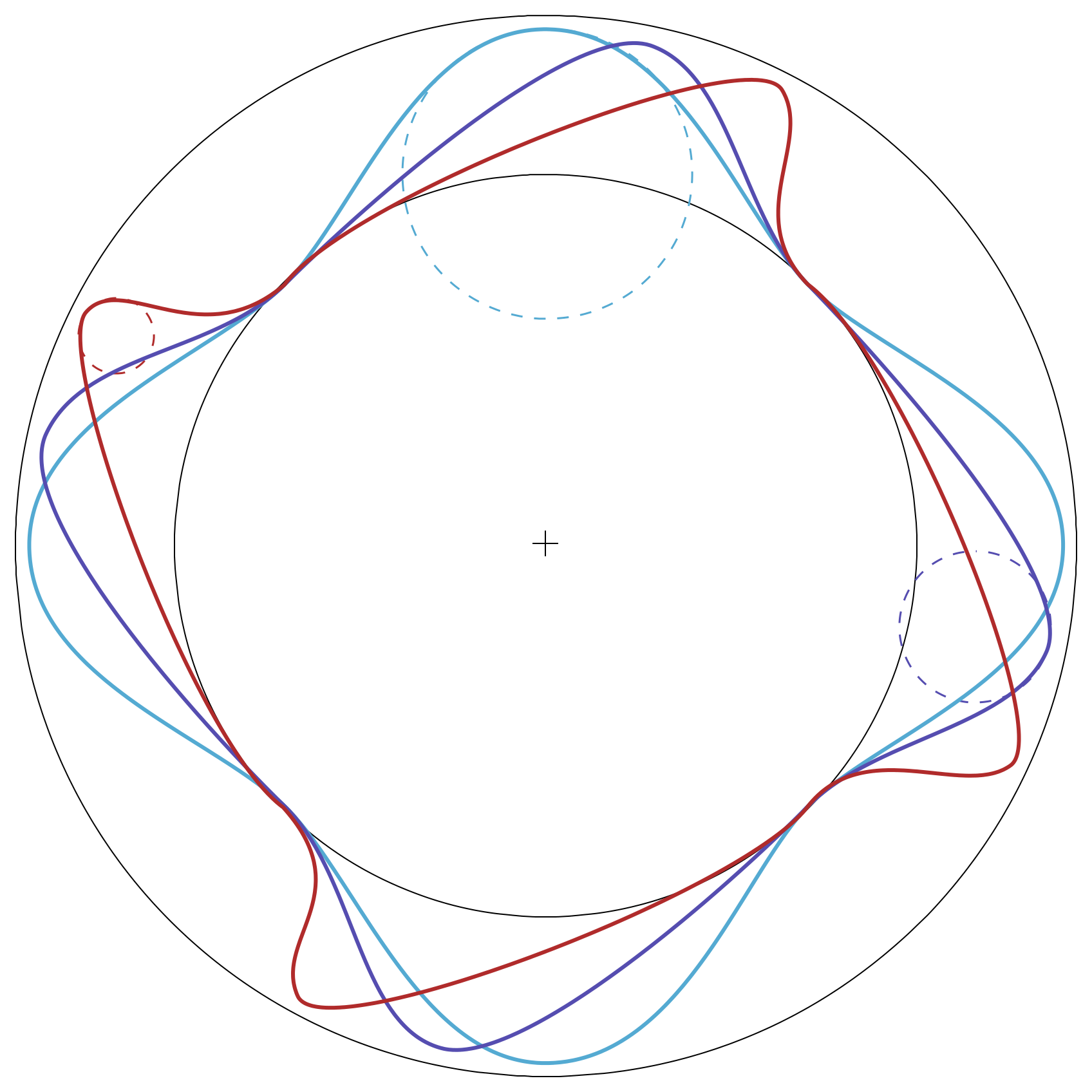}
\caption{\label{fig:sketch-curv-radius} Sketch representing the
         approximate shape of rising flux tubes depending on the
         regime of their rise. The light blue flux tube is in a 
         buoyancy dominated regime. The dark blue flux tube is
         in a transitional regime, where 
         $F_{\rm buoy} \approx F_{\rm corio}$. The red flux tube is in
         a rotation dominated regime.}
\end{figure}

As shown in Fig.~\ref{fig:sketch-curv-radius}, ${\cal R}$ does not only depend on the
azimuthal wavenumber $m$ (4 in that case), but also on the nature of the regime
of the rise. Flux tubes that rise in a {rotation-dominated} regime (red line)
have a smaller curvature radius than flux tubes rising in a buoyancy dominated regime
(blue line). 

It becomes clear that ${\cal R}$ depends on $\beta$, ${\cal M}_{\rm rot}${, } and $m$,
and will here be modeled with the ansatz
%
\begin{equation}
  {\cal R} ~=~ \varpi 
               \beta             ^{f_1} 
               {\cal M}_{\rm rot}^{f_2} ,
\end{equation}
%
where $f_1$ and $f_2$ are two functions of the azimuthal wavenumber
$m$. We can already constrain these functions.
For instance, increasing $\beta$ or ${\cal M}_{\rm rot}$ pushes the
system toward the {rotation-dominated} regime (toward the red
line in Fig.~\ref{fig:sketch-curv-radius}), and decreases 
${\cal R}$. Hence $f_1$ and $f_2$ {both} have to be negative. A further
constraint concerns the axisymmetric case, where the curvature radius
is a constant and equals $\varpi$. In that specific case, 
${\cal R}$ depends neither on $\beta$ nor on ${\cal M}_{\rm rot}$, so
$f_1$ and $f_2$ are both zero. 
%

  In the axisymmetric case, ${\cal R} = \varpi$ and
  the rise velocity $v_{\rm rise}$ is the buoyant velocity $v_{\rm buoy}$.
  In the non-axisymmetric case, the rise velocity is reduced by the tension force.
  We make the hypothesis that the rise velocity is a fraction of the buoyant velocity
  where the fraction is controlled by the curvature radius,
%
\begin{equation}
  \label{eq:rat-v-eq-rat-r}
  v_{\rm rise} ~=~ k \frac{ {\cal R}}{R_{\star}} ~ v_{\rm buoy},
\end{equation}
%
where $v_{\rm buoy}$ is the rise velocity of a flux tube buoyantly rising
in an axisymmetric manner{, and } ${\cal R}$ {is} the
curvature radius of the flux
tube at the apex.
We introduce a factor $k$ to consider the various effects of the
drag force acting on the flux tube, thermal conduction and the
twist. The only constraint on $k$ is that it neither depends on
$\beta$ nor on ${\cal M}_{\rm rot}$. 

  The Coriolis effect depends on the rise velocity{; }
  the reduction of $v_{\rm rise}$ by the magnetic tension naturally alters
  the Coriolis effect:
%
\begin{equation}
  \label{eq:mod-corio}
  F^{\text{*}}_{\rm corio} = 2 \rho ~ v_{\rm buoy} ~ \Omega ~ \frac{{\cal R}}{R_{\star}} {\rm ,}
\end{equation}
%
where $F^{\text{*}}_{\rm corio}$ is the altered Coriolis effect (with respect
to the axisymmetric case).
For the same $\beta$ and ${\cal M}_{\rm rot}$, the Coriolis effect on
an axisymmetric flux tube is going to be larger than on a non-axisymmetric
tube, {simply} because of the reduction of ${\cal R}$ for higher azimuthal
wavenumber.

The regime-controlling relation for the non-axisymmetric case becomes:  
%
\begin{equation}
  \label{eq:fbuoy-o-fcorio-gen}
  \frac{F_{\rm buoy}}{F^{\text{*}}_{\rm corio}} ~=~ \frac{1}{k} ~
              \frac{\Delta \rho ~ g}{2\rho ~ v_{\rm buoy} ~ \Omega} ~
              \frac{R_{\star}}{{\cal R}}.
\end{equation}

Replacing ${\cal R}$, we can rewrite Eq.~(\ref{eq:fbuoy-o-fcorio-gen}) as a
function of $\beta$, ${\cal M}_{\rm rot}$ and $m$:
%
\begin{equation}
  \label{eq:fratio-dimless}
  \frac{F_{\rm buoy}}{F^{\text{*}}_{\rm corio}} ~=~ \frac{{\cal M}_{\rm ff} }{2\,k\,\sqrt{2}} ~
                       \left(\frac{1}{{\cal M}_{\rm rot}}\right)^{1+f_2}
                       \left(\frac{1}{\beta}\right)^{\frac{1}{2}(1+2f_1)}.
\end{equation}
%
The regime of the rise is more rotation dominated for lower azimuthal 
wavenumbers $m$ for given $\beta$ and ${\cal M}_{\rm rot}$.

We can now introduce $\Gamma^{\alpha_2}_{\alpha_1}$ as
%
\begin{equation}
  \Gamma^{\alpha_2}_{\alpha_1} ~=~
                                   \left( \frac{ v_{\rm A}^{\alpha_1} 
                                                 c_{\rm s}^{1 - \alpha_1} } 
                                               { \varpi \Omega }
                                   \right)^{\alpha_2},
\end{equation}
%
with 
%
\begin{eqnarray}
  \label{eq:alphas-def}
  \alpha_{1} ~&=~ \frac{1+2f_1}{1+f_2} ~~\rm{,} \nonumber\\ \nonumber\\
  \alpha_{2} ~&=~ 1+f_2 ~~\rm{.}
\end{eqnarray}
%
$\Gamma_{\alpha_1}^{\alpha_2}$ can be seen as a modified ${\rm Lo}$ that
compares the ratio of buoyant force over altered Coriolis effect.
Because ${\cal M}_{\rm ff}$, and $k$ are constant for a given
$m$ and both independent of $\beta$ and ${\cal M}_{\rm rot}$, we can
introduce $\Gamma_{\alpha_1}^{\alpha_2}$ in
Eq.~(\ref{eq:fratio-dimless}) and demonstrate that the latter acts as a
proxy to the force ratio
%
\begin{equation}
  \label{eq:fratio-propto-gamma}
  \frac{F_{\rm buoy}}{F^{\text{*}}_{\rm corio}} = 
      \frac{{\cal M}_{\rm ff}}{2\,k\,\sqrt{2}} 
      \left(
          \frac{\gamma}{2}
      \right)^{\frac{\alpha_1 \alpha_2}{2}} 
~~\Gamma_{\alpha_1}^{\alpha_2}.
\end{equation}
%
It can be {seen} that in the axisymmetric case, where $f_1$ and $f_2$ are
zero, $\alpha_1$ and $\alpha_2$ both become unity and
$\Gamma_{\alpha_1}^{\alpha_2}$ becomes 
%
\begin{equation}
  \Gamma^{1}_{1} ~=~ \frac{ v_{\rm A}     }
                          { \varpi \Omega }
                 ~\equiv~ {\rm Lo}
                 ~\propto~ \frac{F_{\rm buoy}}{F_{\rm corio}} ~~\rm{.}
\end{equation}
%
The definition of $\Gamma^{1}_{1}$ recovers the axisymmetric ($m=0$)
results. 

To conclude this section, we identified 
a dimensionless number $\Gamma^{\alpha_2}_{\alpha_1}$, that acts as a 
proxy to the force ratio. As such $\Gamma_{\alpha_1}^{\alpha_2}$
controls the regime of the rise for $m=0$ and $m>0$.

Assuming a given azimuthal wavenumber $m$ for the initial conditions,
we predict that two simulations with the same
$\Gamma_{\alpha_1}^{\alpha_2}$ will reveal the same solution. As a
result, the relative {rise-time} of a magnetic flux tube in a rotating
stellar interior should scale with $\Gamma_{\alpha_1}^{\alpha_2}$. In
the following sections  
(\ref{sec:validation-of-the-setup-with-a-2d-numerical-experiment} and
\ref{sec:extension-to-the-non-axisymmetric-case-3d}), we will verify
this prediction by studying the behavior of two series of numerical
experiments.

\section{Validation of the setup with a 2D numerical experiment}
\label{sec:validation-of-the-setup-with-a-2d-numerical-experiment}

In agreement with
\cite{Schussler+92}, we have shown that for the axisymmetric case 
$\Gamma_{\alpha_1}^{\alpha_2}$ reduces to the Lorentz number
($\Gamma^1_1$). We already know that under   
the {thin-flux-tube} approximation the relative {rise-time} scales with
$\Gamma^1_1$ \citep{Choudhuri+87, Schussler+92}. 
Hence, we are interested in verifying whether this scaling behavior holds 
in the case of compressible simulations. For this purpose we carried
out a series of simulations based on the STD-2D setup.

\subsection{Parameter study}
\label{sec:4--carryiing-a-parameter-study}

\begin{figure}[bpth!]
  \centering
  \includegraphics[width=1.0\linewidth, trim=10 35 -15 15, clip]{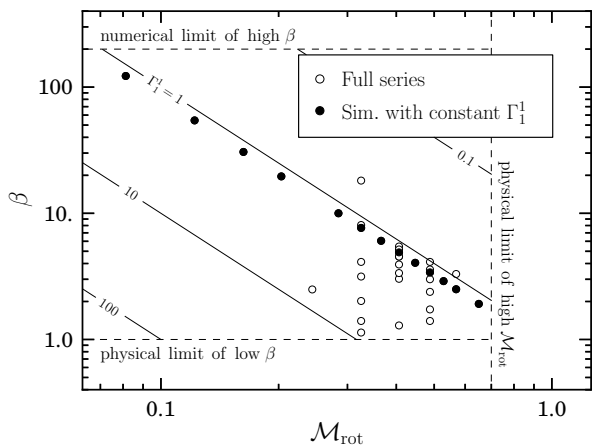}
  \caption{\label{fig:beta-o-mrot-2D} Parameter space of the 2D
           series, where each point represents one simulation. The
           filled symbols represent simulations for which
           $\Gamma_{1}^{1}$ is a constant. 
           Regions that are unphysical or numerically 
           inaccessible are delimited by the dashed lines. 
           The low-$\beta$ as well as
           the high-$\cal{M}_{\rm{rot}}$ limits are due to physical
           constraints. 
           The high-$\beta$ limit is of numerical nature. 
          } 
\end{figure}

The main goal of this section is to study the effect of rotation on
the axisymmetric rise of a magnetic flux tube. This effect is
controlled by $\Gamma^1_1$. The latter is a composition of
$\cal{M}_{\rm rot}$ and the plasma-$\beta$, with $\cal{M}_{\rm rot}$ 
defining the rotational velocity of the star and the plasma-$\beta$
the buoyancy of the flux tube. In the case of an isothermal flux tube, 
$\beta$ is directly proportional to the lack of density inside the
flux tube, {that is, } the strength of the buoyant force. Hence, the
parameter study is carried out in the ($\beta$, $\cal{M}_{\rm{rot}}$)
parameter space. As seen in Fig.~\ref{fig:beta-o-mrot-2D},
we covered two orders of magnitude for the $\beta$ parameter and about 
one order of magnitude for the ${\cal M}_{\rm rot}$ parameter. This
large domain is restricted by physical and numerical limits. Beyond
those limits, simulations deliver either unreliable results due to
high numerical diffusion ($\beta > 200$), or results that are not
applicable to stellar interiors, such as  $\beta < 1$ or rotation
velocities being too close to the sound speed (${\cal M}_{\rm rot}
\approx 1$).

\subsection{Verification of the scaling behavior}
\label{sec:4--verifying-the-scaling-behavior}

\begin{figure}[pbth!]
  \centering
  \includegraphics[width=1.0\linewidth, trim=10 35 -15 15, clip]{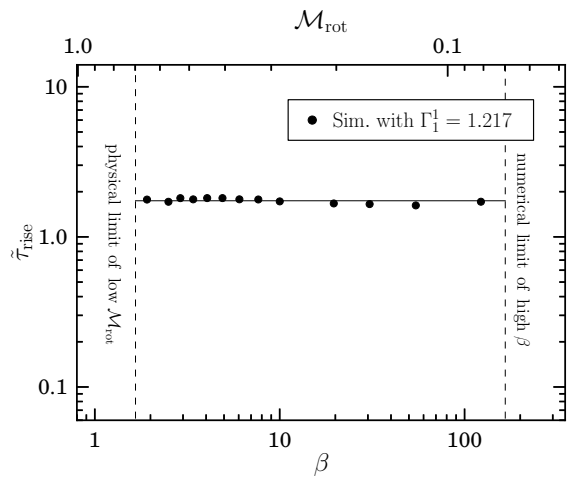}
  \caption{\label{fig:scaling-2D} Relative rise-time
           $\tilde{\tau}_{\rm{rise}}$ versus plasma-$\beta$
           also showing the corresponding
           $\cal{M}_{\rm{rot}}$ for a constant $\Gamma_{1}^{1}$. Each
           point represents a simulation of the STD-2D series where we
           vary $\beta$ from about 1 to 100. 
          } 
\end{figure}

\begin{figure*}[bpth!]
\centering
  \begin{tabular}{ll}
    \includegraphics[width=.5\linewidth, trim=130 130 100 100, clip]{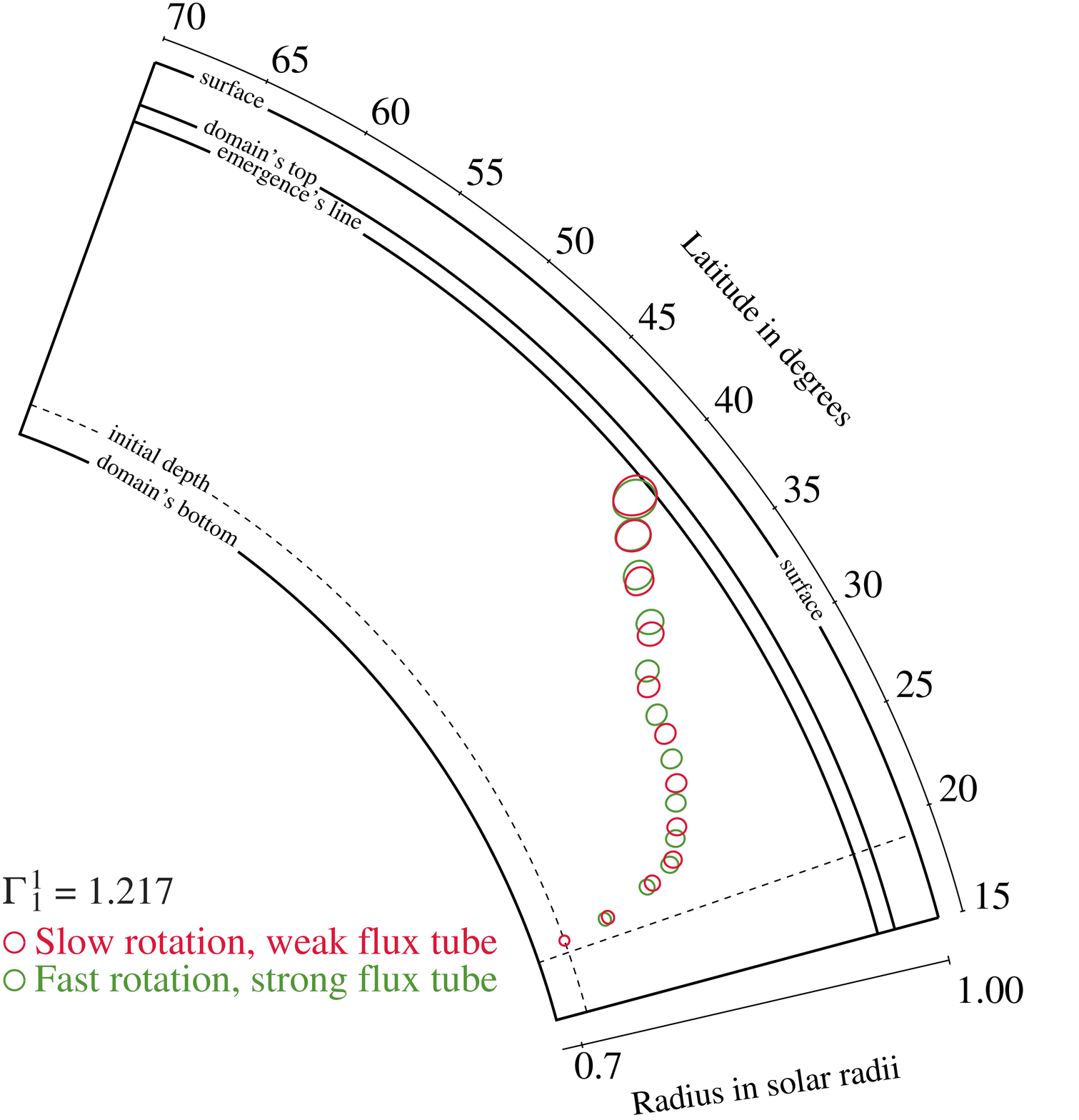} &
    \includegraphics[width=.5\linewidth, trim=130 130 100 100, clip]{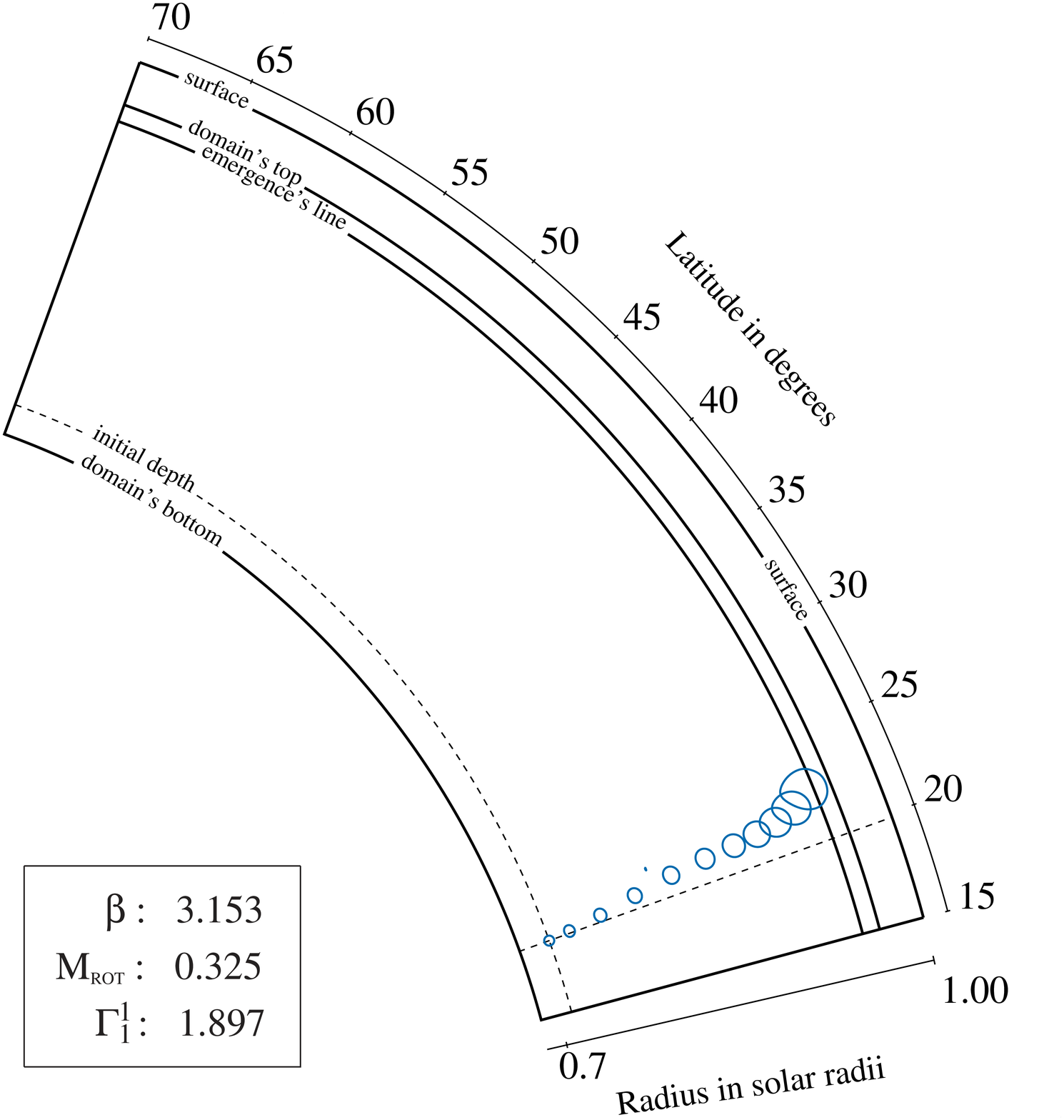} \\
    (a) &     (b) \\ & \\ & \\
    \includegraphics[width=.5\linewidth, trim=130 130 100 100, clip]{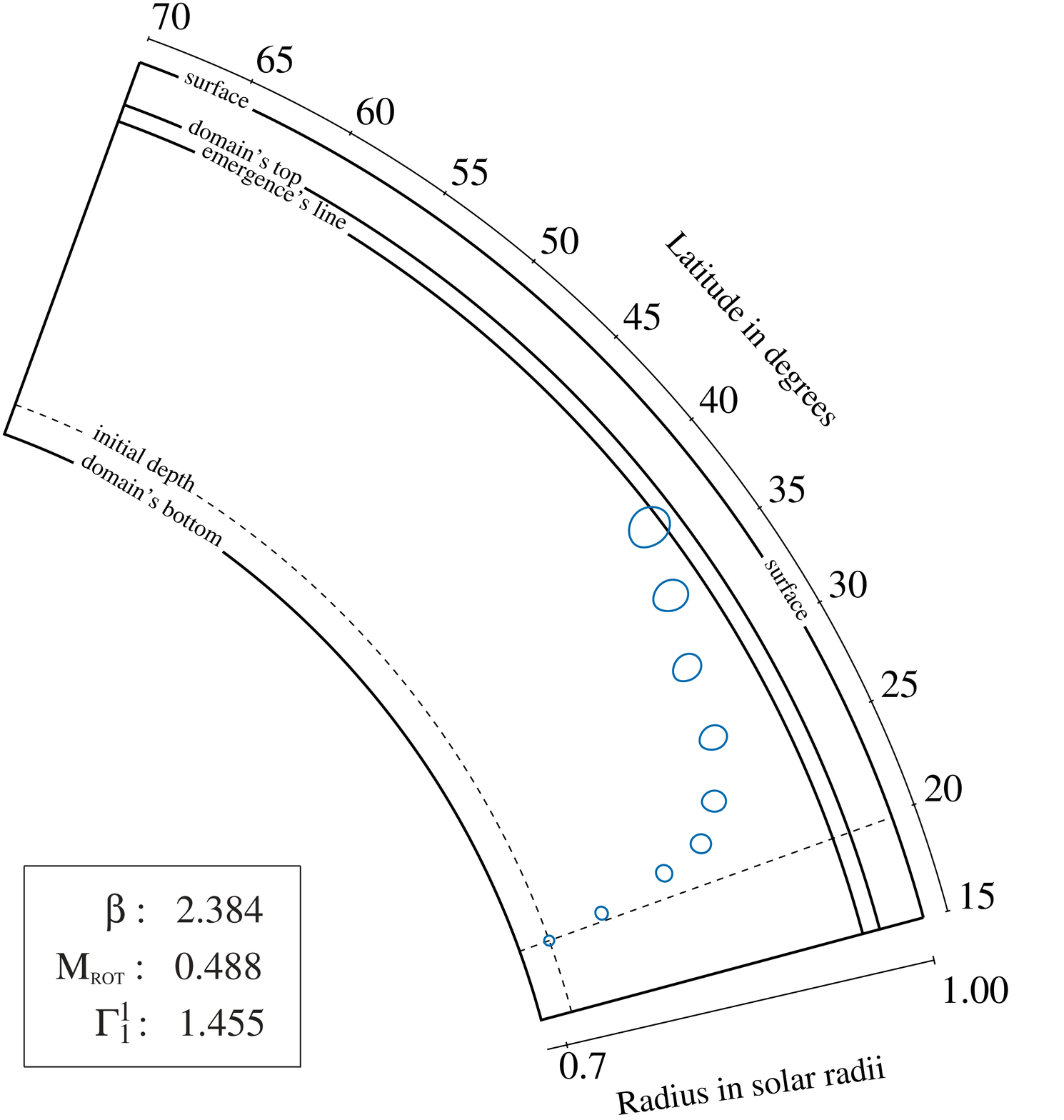} &
    \includegraphics[width=.5\linewidth, trim=130 130 100 100, clip]{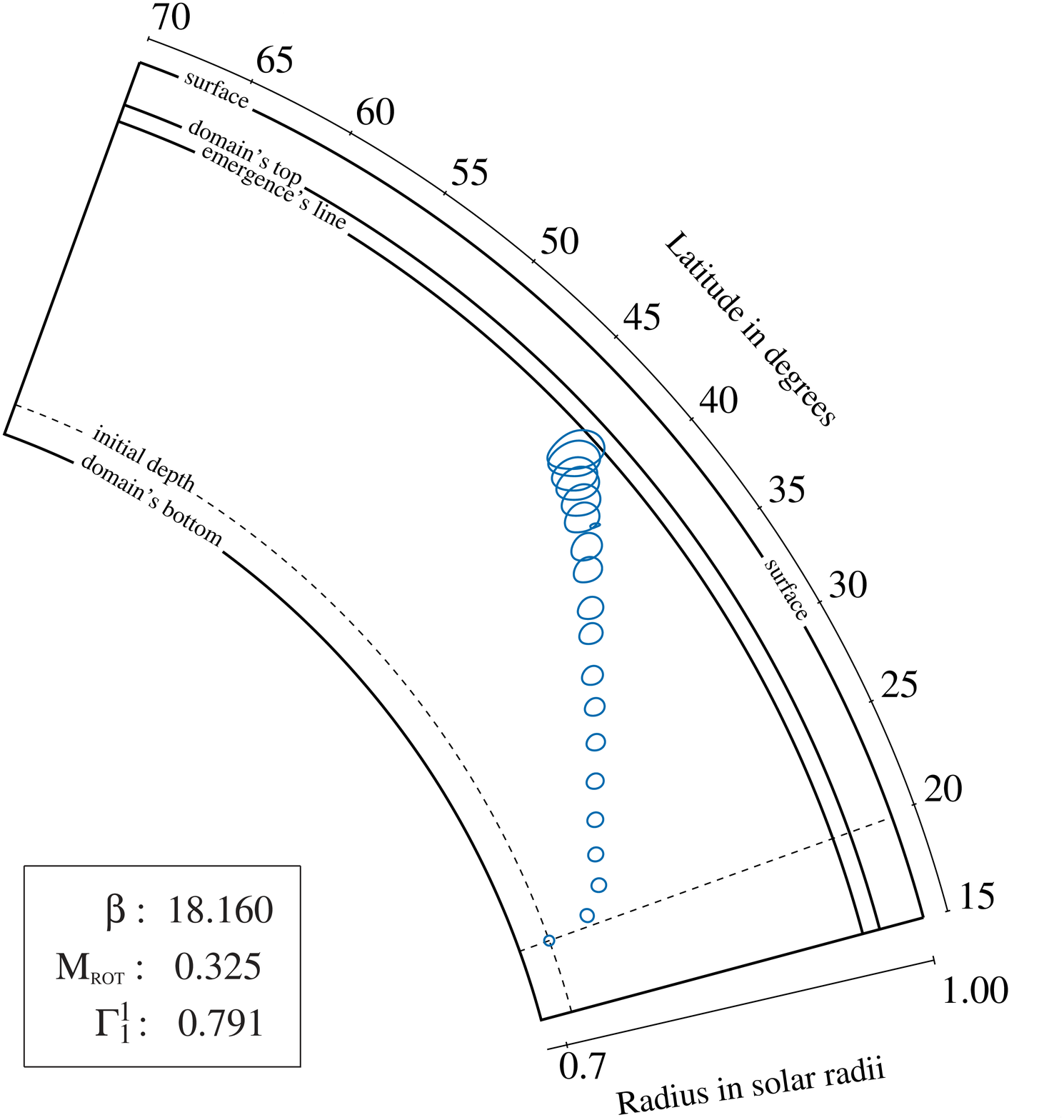}   \\
    (c) &     (d) \\
  \end{tabular}
  \caption{\label{fig:paths} Cross-sections of the magnetic field at various
           times for five representative simulations, representing the
           path taken by the flux tube all along its rise. (a) is a
           plot of two simulations with the same $\Gamma_{1}^{1}$
           (same regime): the green contours correspond to a strongly
           buoyant flux tube in a rapidly rotating interior (labeled
           a.1 in Table~\ref{tab:list-of-sim2D}), {and} the red contours
           correspond to a weakly buoyant flux tube in a slowly
           rotating interior (labeled a.2 in
           Table~\ref{tab:list-of-sim2D}). Panels (b), (c), and (d)
           show the dependence of the path on $\Gamma_{1}^{1}$, in three different
           simulations.  
          } 
\end{figure*}

From a derivation of the force balance, we predicted in
Sect.~\ref{sec:scaling-relation-of-rising-flux-tube} that scalable
properties of the rise of a magnetic flux tube should scale with 
$\Gamma^1_1$. In order to verify our prediction we selected a set of
simulations out of the parameter study described above. This subset
corresponds to the filled symbols in
Fig.~\ref{fig:beta-o-mrot-2D}. In this sub-series we fixed
$\Gamma^1_1$ and evolved ${\cal M}_{\rm{rot}}$ as well as the
plasma-$\beta$ accordingly. The results are demonstrated in two
{Figures}. First, Fig.~\ref{fig:scaling-2D} shows that the relative rise
time is constant for a constant $\Gamma^1_1$ ($=1.217$), independently
of $\beta$ and ${\cal M}_{\rm{rot}}$. Second, the top-left panel (a)
of Fig.~\ref{fig:paths} shows the path of two flux tubes with the same 
$\Gamma^1_1$ but different values of ${\cal M}_{\rm{rot}}$ and
$\beta$. From these plots, we find that over two orders of magnitude
in $\beta${, } neither the {rise-time} nor the path of the flux tube
changed significantly. This is {a} strong evidence that 
our axisymmetric setup {scales} with $\Gamma^1_1${, } as predicted by
both the {thin-flux-tube} approximation and our theoretical derivation
(Sect.~\ref{sec:scaling-relation-of-rising-flux-tube}). The good 
scaling behavior convinced us that our setup delivers results comparable
to hypothetical simulations with $\beta ~=~ 10^{5}$. It
demonstrates the possibility of computing compressible simulations of
magnetic flux tubes that rise in the same regime as in the Sun.  
%

%

\subsection{Validation by morphological study}
\label{sec:4--validation-by-morphological-study}

In this section we address the question of the influence of
$\Gamma^1_1$ on the path taken by the flux tube while rising. This
question has been extensively studied 
\citep{Choudhuri+87,Choudhuri+90,Schussler+92,Caligari+95,Caligari+96,Deluca+97,Fan08}.  
We now show that our compressible setup reproduces the behavior
found in the literature. In the axisymmetric case, the flux tube
remains a torus {throughout} its rise. Furthermore, since our setup does
not take into account turbulent convection, the angular momentum of
the flux tube remains {almost} constant. We assume that molecular diffusion
is not sufficient to transfer angular momentum from the flux tube to
its surroundings. Therefore, an infinitely slowly rising flux tube
will follow the path of constant angular momentum on which it lies 
initially. At finite rise speed, the flux tube follows a more complex
path. This is illustrated in the panels (b), (c) and (d) of
Fig.~\ref{fig:paths}. For a high $\Gamma^1_1$ (b) the flux tube rises
radially; the flux tube is in a {buoyancy-dominated} regime. For
a $\Gamma^1_1$ {of approximately} unity (c) the flux tube rises first radially and
then axially{; } in that case the buoyant force and the Coriolis force
acting on the flux tube are of the same order {and} the flux tube is in a
transitional regime. A flux tube in a 
rotation dominated regime (d) will rise {almost} axially. These
behaviors have been shown to be exclusively axisymmetric
\citep{Fan08}. Since our setup is able to reproduce results from the
literature -- {thin-flux-tube} results \citep{Schussler+92} as well as
anelastic results \citep{Fan08} -- for corresponding values of
$\Gamma^1_1$, we are confident that our axisymmetric setup is
reliable. It also shows that compressibility does not influence the
path of the flux tube within the computational domain
($r<0.964R_\star$).

\subsection{General relation for the relative {rise-time} in 2D}
\label{sec:4--extracting-a-general-relation-for-the-relative-rise-time}

\begin{figure}[bpth!]
  \centering
  \includegraphics[width=1.0\linewidth, trim=15 20 15 19, clip]{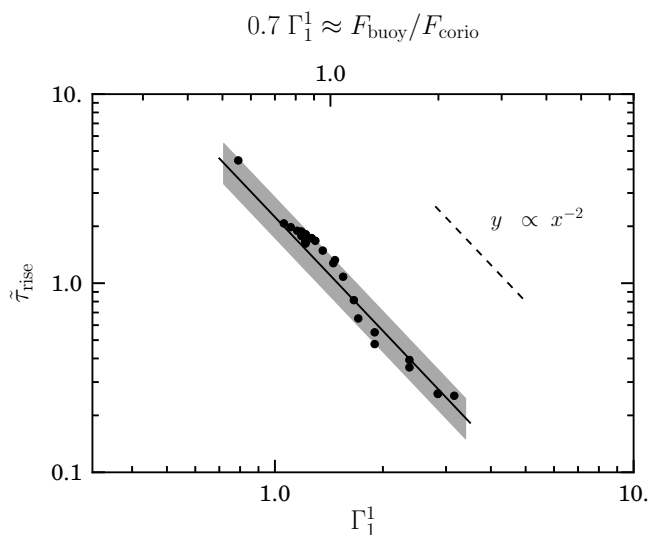}
  \caption{\label{fig:rrt-o-rom-2D-dumfit} Relative {rise-time} 
           $\tilde{\tau}_{\rm{rise}}$ versus the scaling parameter  
           $\Gamma^{1}_{1}$ for the STD-2D series. 
           We obtain a self-similar function of power $-2$. The {gray}
           zone illustrates the low scatter of the individual
           values. The upper axis indicates the corresponding values
           of $0.7\,\Gamma_{1}^{1}$ which represents the best estimate for 
           $F_{\rm buoy}/F_{\rm corio}$ we could extract from our 
           simulations. 
           } 
\end{figure}

By carrying out a set of numerical experiments based on the STD-2D
setup, varying only $\Gamma^1_1$, we investigated the influence of
the {rotation-dominated or buoyancy-dominated} character
of the regime on the relative rise-time. If the dependence is
approximated by a power law as shown in
Fig.~\ref{fig:rrt-o-rom-2D-dumfit}, we obtain a relation 
%
\begin{equation}
  \label{eq:scaling-rel-2D}
  \tilde{\tau}_{\rm{rise}} = 2.24~ \left(\Gamma^1_1\right)^{-2}.
\end{equation}
%
Unfortunately the proportionality factor relating $\Gamma_{1}^{1}$ and
$F_{\rm buoy}/F_{\rm corio}$ in Eq.~(\ref{eq:fratio-propto-gamma})
remains unknown and Eq.~(\ref{eq:scaling-rel-2D}) fails to identify
the nature of the regime. Several series would be required to fully 
identify how the proportionality factor depends on the drag force, the
twist and thermal conduction. Instead, making use of a 
morphologic argument one can estimate the proportionality factor: 
in panel (c) of Fig.~\ref{fig:paths} the magnetic flux tube starts
its latitudinal deflection at half of the convective zone; this
indicates that for $\Gamma_{1}^{1} \approx 1.5$ the magnetic flux tube
rises in a transitional regime. We assume $F_{\rm buoy}/F_{\rm corio} \approx 1$ 
for this case and estimate the proportionality factor to be about $0.7$. In
Fig.~\ref{fig:rrt-o-rom-2D-dumfit} we indicate the corresponding
estimate of $F_{\rm buoy}/F_{\rm corio}$ on the upper
axis. Simulations lying on the~left-hand side of unity are in a 
{rotation-dominated} regime, whereas simulations on
the~right-hand side of unity reveal a {buoyancy-dominated} regime.
%
%
The most relevant information to extract from 
Fig.~\ref{fig:rrt-o-rom-2D-dumfit} is the power of $-2$ of the
self-similar function that relates the relative rise 
time with $\Gamma_{1}^{1}$. We call this exponent $\alpha_3$. 
%
\begin{equation}
  \tilde{\tau}_{\rm{rise}} 
  \propto \left( \Gamma_{\alpha_1}^{\alpha_2} \right)^{\alpha_3} ~{\rm ,}
\end{equation}
%
with $\alpha_1 = \alpha_2 = 1$ and $\alpha_3 = -2$, for axisymmetric rises.
This result agrees with \citet{Jouve+10}:
for a given rotation rate the relative {rise-time} is proportional to
the inverse square of the magnetic field strength of the flux tube, {that is,}
%
\begin{equation}
  \tilde{\tau}_{\rm{rise}} 
  \propto \left(\frac{B_{\phi}}{B_{\rm eq}} \right)^{-2}.
\end{equation}
%
We emphasize that this result also agrees with \citet{Schussler+92}
who underlined the fact that the {rise-time} is a good scaling parameter
for the present problem. However, the {rise-time} is an 
a posteriori quantity. They showed therefore that the ratio of 
buoyancy over Coriolis force is a reasonable proxy for the rise-time,
and the {rise-time} can be used as a scaling parameter. In the present
work, we choose $\Gamma^1_1$ which, being an a priori quantity,
is a more appropriate scaling parameter for the physical problem
considered. Nevertheless, their conclusions are compatible with
ours. Such an agreement shows that the {thin-flux-tube} approximation
is a good approximation in 2D.

\section{Extension to the non-axisymmetric case (3D)}
\label{sec:extension-to-the-non-axisymmetric-case-3d}

In Sect.~\ref{sec:scaling-relation-of-rising-flux-tube}, we have
predicted that in the non-axisymmetric case{, } the regime of the rise 
of the flux tube is not controlled by $\Gamma^1_1$ anymore, but by a
$\Gamma_{\alpha_1}^{\alpha_2}$, with $\alpha_1$ and $\alpha_2$ being
less than unity. In this section we compute both $\alpha_1$ and
$\alpha_2$ from measurements of the curvature radius ${\cal R}$ of
magnetic flux tubes; we verify that the setup scales with the
resulting dimensionless parameter; and confirm the predictions.

\subsection{Defining the parameter study}
\label{sec:5--defining-the-parameter-study}

\begin{figure}[bpht!]
  \centering
  \includegraphics[width=1.0\linewidth, trim=10 35 -15 15, clip]{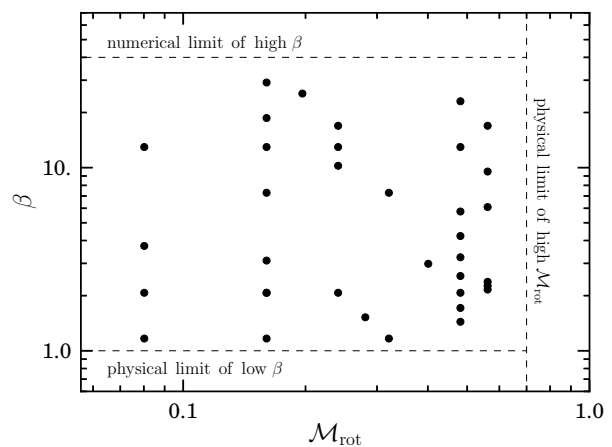}
  \caption{\label{fig:beta-o-mrot-3D} The ($\beta$, 
           ${\cal M}_{\rm rot}$) parameter space of the STD-3D series,
           for an azimuthal wavenumber $m = 8$. Each point represents a
           simulation. Regions that are unphysical or numerically 
           inaccessible are delimited by the dash lines. 
           In a similar manner as for the STD-2D series, the
           low-$\beta$ and high-${\cal M}_{\rm rot}$ limits lead to
           unrealistic physical regimes, while the high-$\beta$ limit
           is numerical. 
           }
\end{figure}

\begin{figure*}[bpht]
  \centering
  \includegraphics[width=0.95\linewidth]{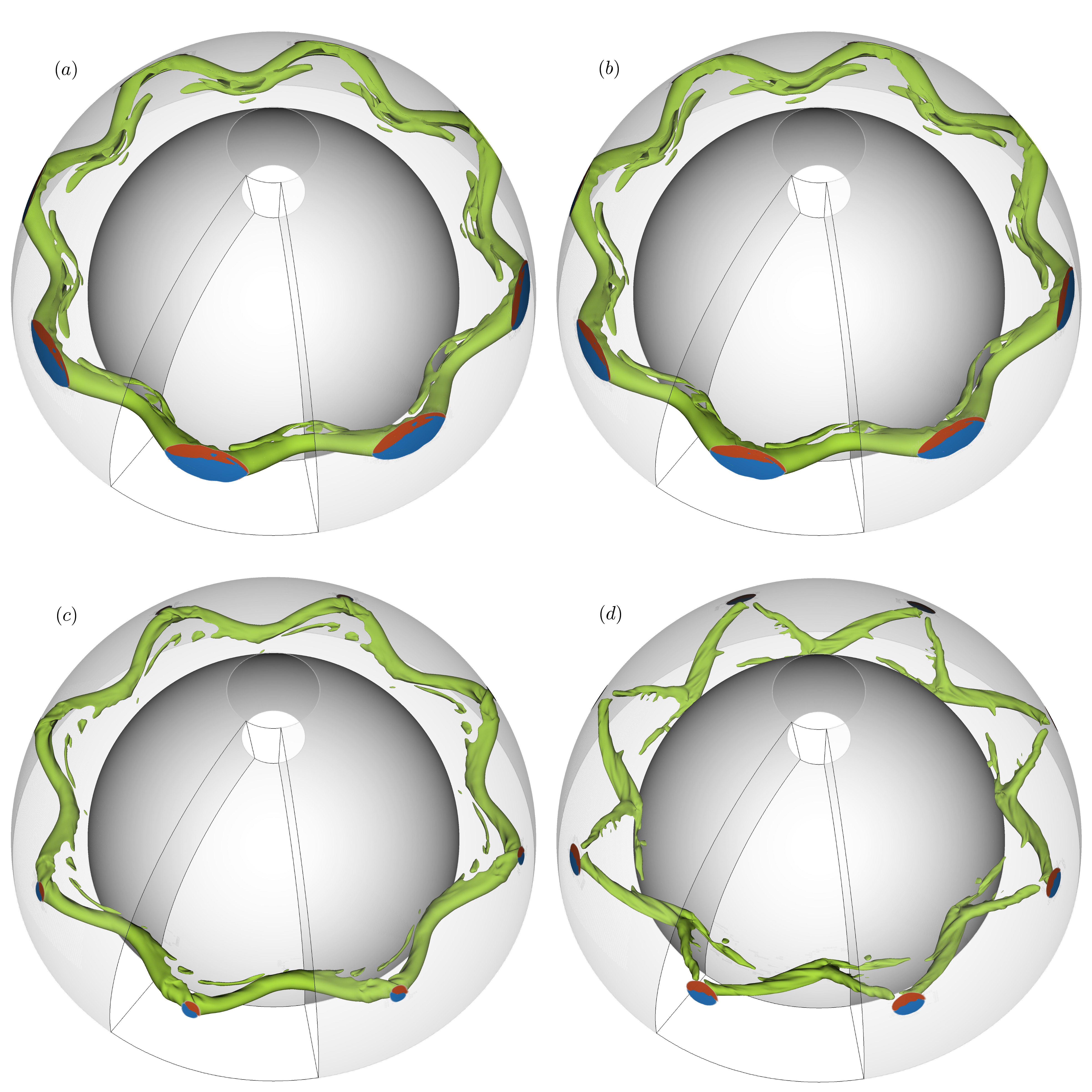}
  \caption{\label{fig:3D-rise} Morphology of the flux tube when it
           emerges for the four different cases: panels (a)--(d). The 
           simulations are ordered with increasing
           $\Gamma^{0.855}_{0.793}$ running from the 
           {buoyancy-dominated} regime (a) through (d) which is
           the {rotation-dominated} regime. 
          }
\end{figure*}

In order to verify the existence of a scaling behavior of the setup,
we conducted a parameter study for the non-axisymmetric case based on
the STD-3D setup. We carried out this study in the same parameter
space ($\beta$, ${\cal M}_{\rm{rot}}$) as for the 2D series. In order
to visualize this study, Fig.~\ref{fig:beta-o-mrot-3D} shows the
positions of all simulations in the parameter plane;
while Fig.~\ref{fig:3D-rise} illustrates the morphology of four
representative flux tubes reaching the surface. We emphasize that,
due to the lower resolution of the STD-3D model, the numerical limit
on $\beta$ decreased from 200 to 40, as compared to the STD-2D case.

%
%

\subsection{Validation of the general scaling parameter}
\label{sec:5--validation-of-the-general-scaling-parameter}

In Sect.~\ref{sec:scaling-relation-of-rising-flux-tube} we
predicted that the regime of the rise is controlled by
$\Gamma^{\alpha_2}_{\alpha_1}$ where $\alpha_1$ and $\alpha_2$ are
functions of $f_1$ and $f_2$. In order to validate this idea, we need
to compute $f_1$ and $f_2$ from the simulations.
  To do so, we first need the curvature radius, ${\cal R}$.

\begin{figure*}[bpht!]
\centering
  \begin{tabular}{ccc}
    \includegraphics[width=.3\linewidth, trim=60 60 60 60, clip]{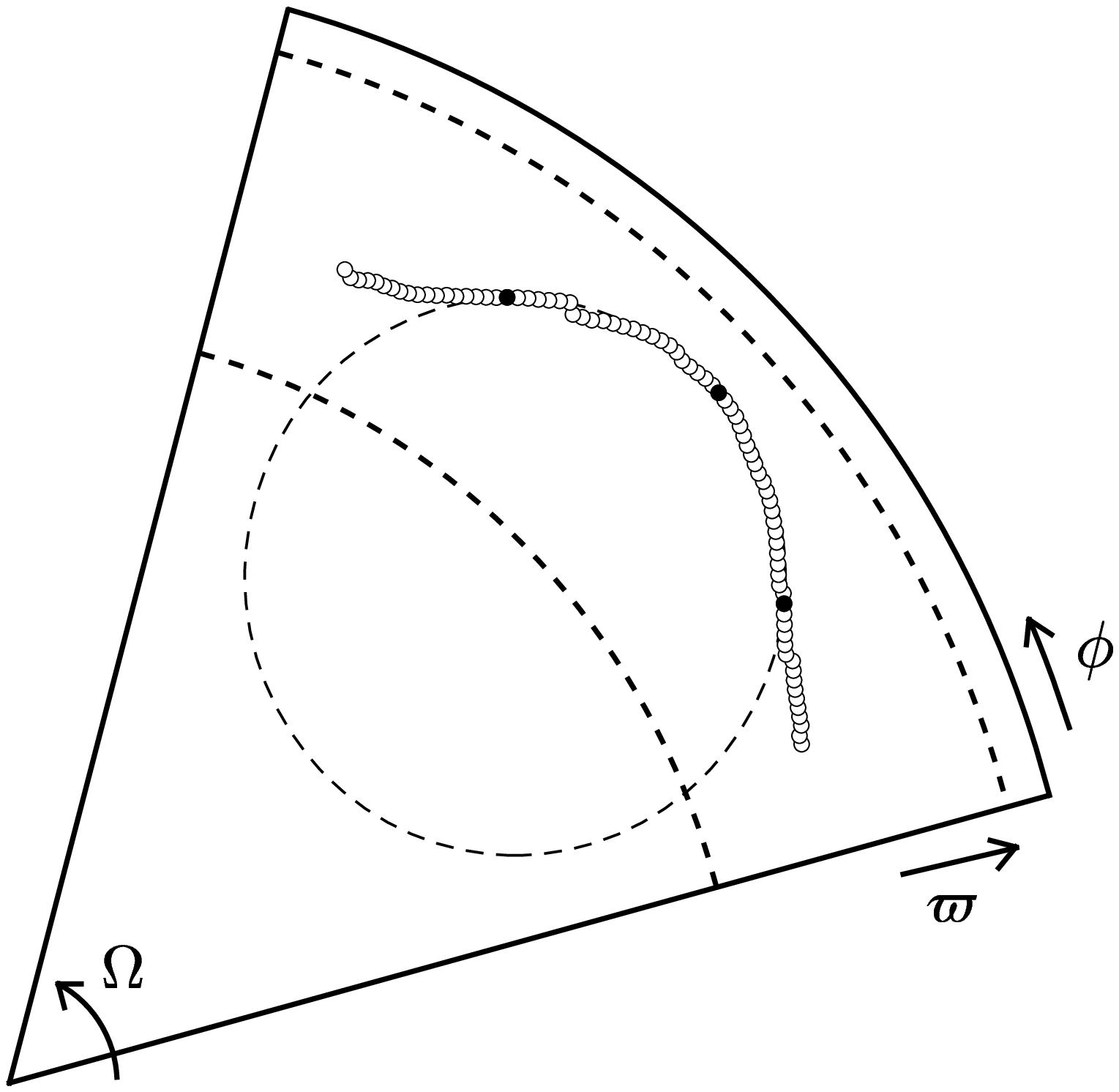} &
    \includegraphics[width=.3\linewidth, trim=60 60 60 60, clip]{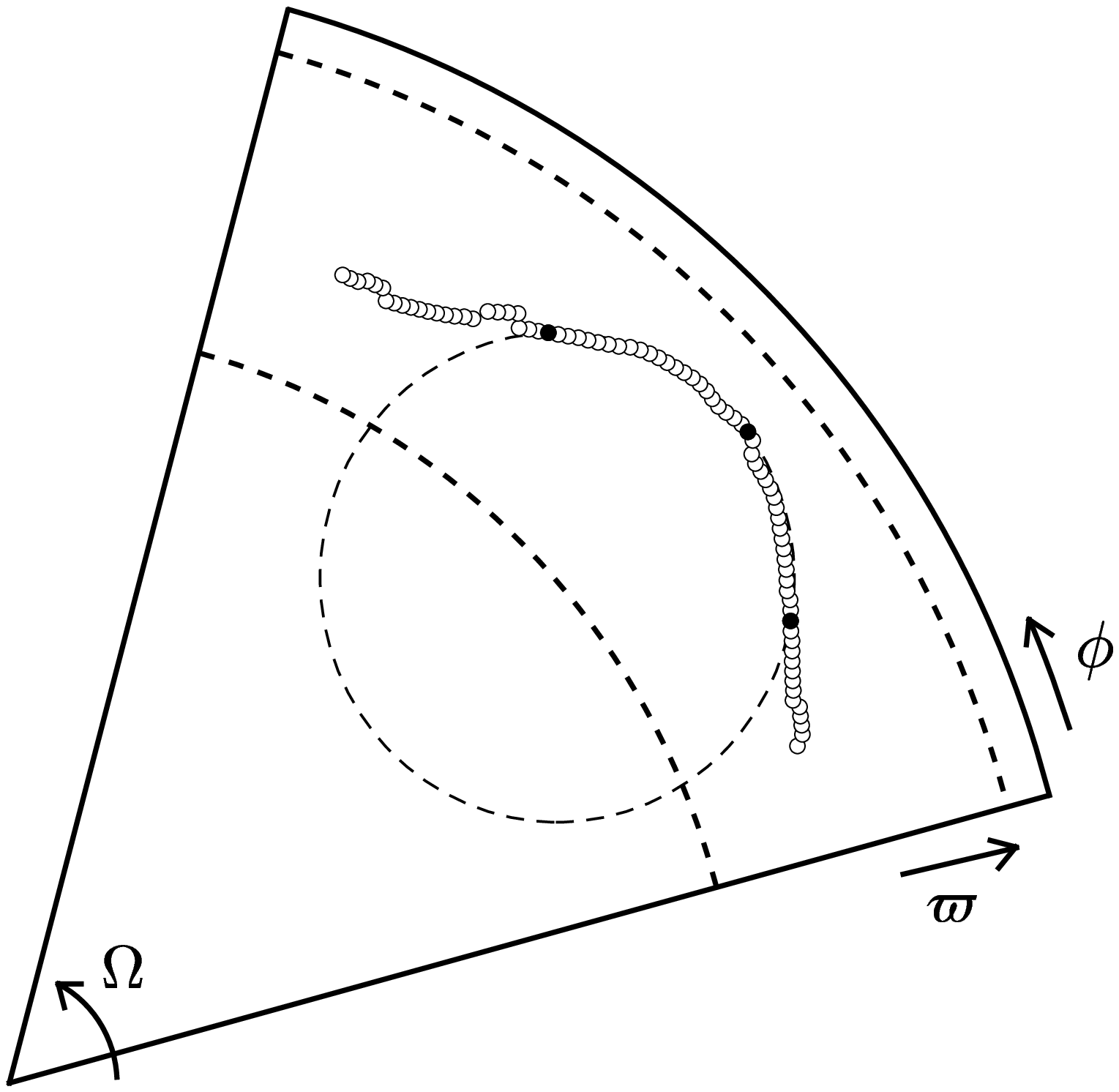} &
    \includegraphics[width=.3\linewidth, trim=60 60 60 60, clip]{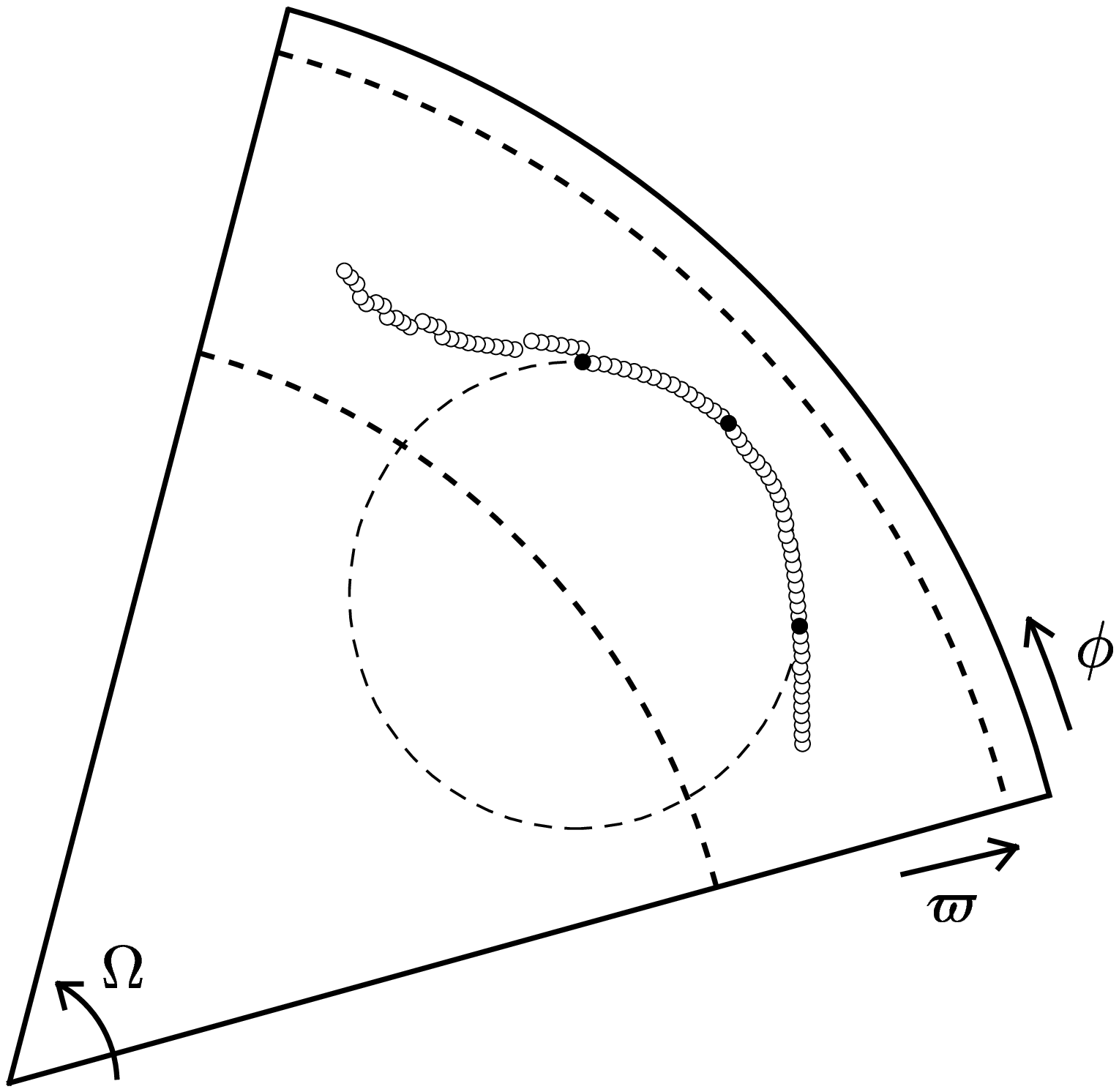} \\[-8.pt]
    $(a)$ &     $(b)$ &    $(c)$ \\ 
        &         &        \\[-4.pt]
    \includegraphics[width=.3\linewidth, trim=60 60 60 60, clip]{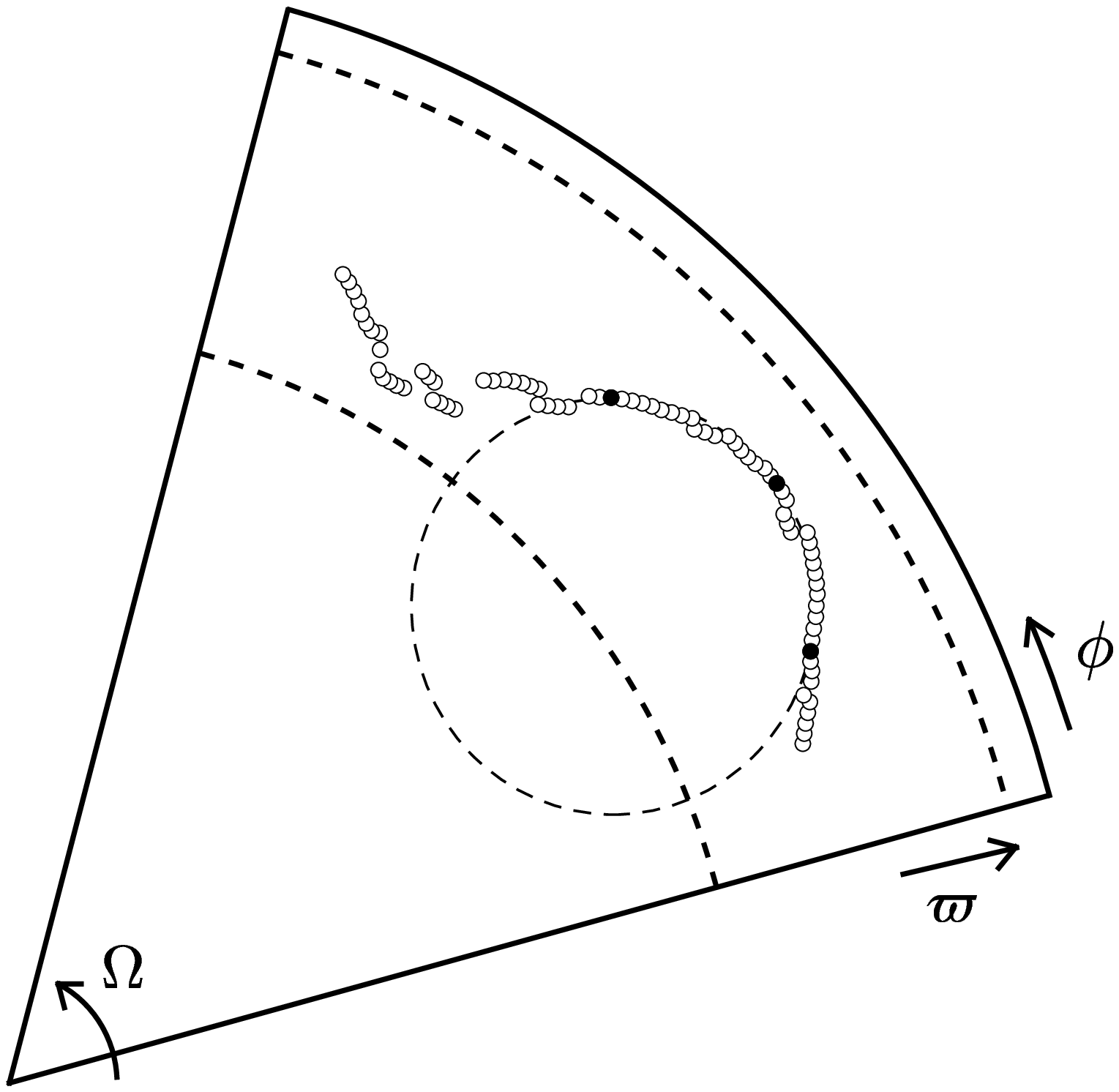} &
    \includegraphics[width=.3\linewidth, trim=60 60 60 60, clip]{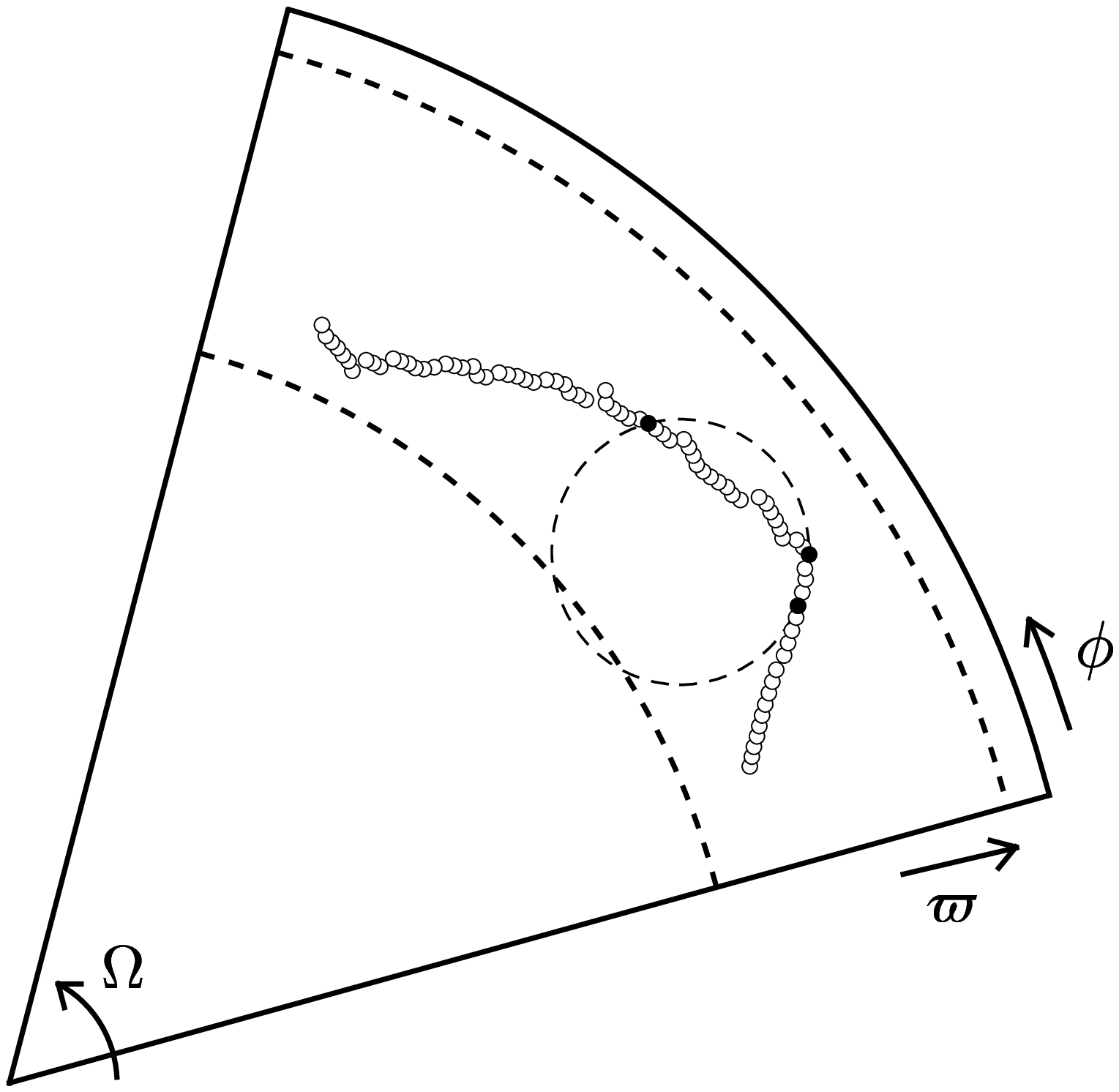} &
    \includegraphics[width=.3\linewidth, trim=60 60 60 60, clip]{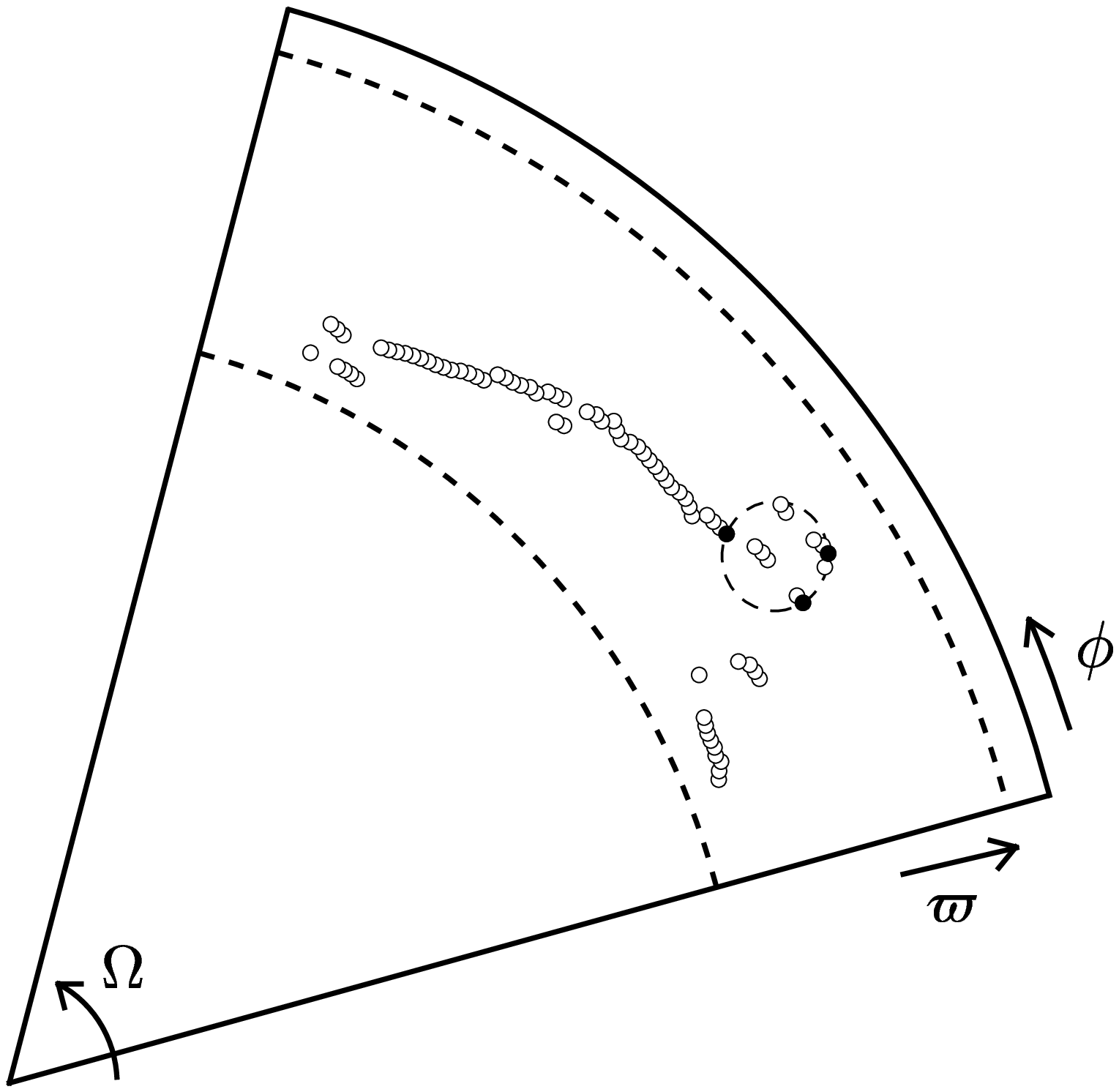} \\[-8.pt]
    $(d)$ &     $(e)$ &    $(f)$ \\
  \end{tabular}
  \caption{\label{fig:curv-radius} Projection of magnetic flux tubes
    on a horizontal plane cutting the northern hemisphere of the Sun at
           $z=0.34~R_{\star}$ (with $z$ being the height in cylindrical
           coordinates). Each point represents the projected position of
           the maximum entropy inside the flux tube in a given
           meridional plane. These points are supposed to map the
           center line of the magnetic flux tube. The radii of the
           dashed circles are a proxy for the curvature radii 
           ${\cal R}$ of the flux tubes at the apex. Panels
           (a)--(d) show suitable estimates of ${\cal R}$. Panels (e)
           and (f) show cases where the estimation fails. 
           }
\end{figure*}

Computing the curvature radius is not a straightforward task.
We use a method that consists in projecting the flux tube on a horizontal plane
cutting the northern hemisphere of the star at a vertical distance
from the equatorial plane of $z=0.34~R_{\star}$. From this projection
we construct the circle going through the apex and two points, one of
each leg of the loop, being two pressure scale heights deeper than the
apex. We consider the radius of such a circle to be a good
approximation for ${\cal R}$. 

In Fig.~\ref{fig:curv-radius} we plot
a few examples of the circles we obtained from this simple method. It is
important to note that in the case when the {rise-time} is larger than
the growth time of the kink instability (all our flux tubes are kink
unstable) the shape of the flux tube becomes more complex than just a
simple $\Omega$-loop. In such cases the simple method we used to determine
  ${\cal R}$ fails; the values for ${\cal R}$ are too large.
Such situations are illustrated in panels ($e$) and ($f$)
of Fig.~\ref{fig:curv-radius}. We exclude
these cases from the computation of $f_1$ and $f_2$ and emphasize
them in Table~\ref{tab:list-3D} in italics. The four other panels 
($a$) -- ($d$) in Fig.~\ref{fig:curv-radius} are ordered by
decreasing ${\cal R}$. This {Figure} shows visually the quality of the
method to extract ${\cal R}$.

\begin{figure}[bpht]
  \centering
  \includegraphics[width=1.0\linewidth, trim=10 10 -15 15, clip]{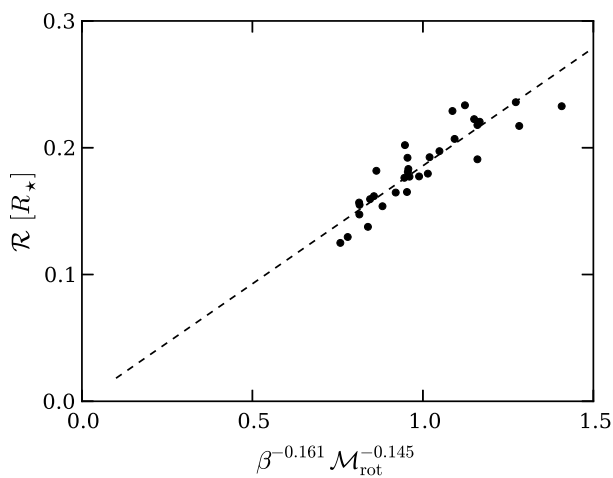}
  \caption{\label{fig:rad-o-beta-mrot} Dependency of the curvature
          radius ${\cal R}$ at the apex, on both $\beta$ and 
          ${\cal M}_{\rm rot}$. We obtain both exponents $f_1$ and
          $f_2$ by minimizing the residual of the best
          fit. \textit{Notes:} not all simulations are present in this
          plot. We excluded the simulations for which the method we
          use to compute the curvature radius gives unreliable results. 
          } 
\end{figure}

\begin{figure}[bpht]
  \centering
  \includegraphics[width=1.0\linewidth, trim=10 10 -15 15, clip]{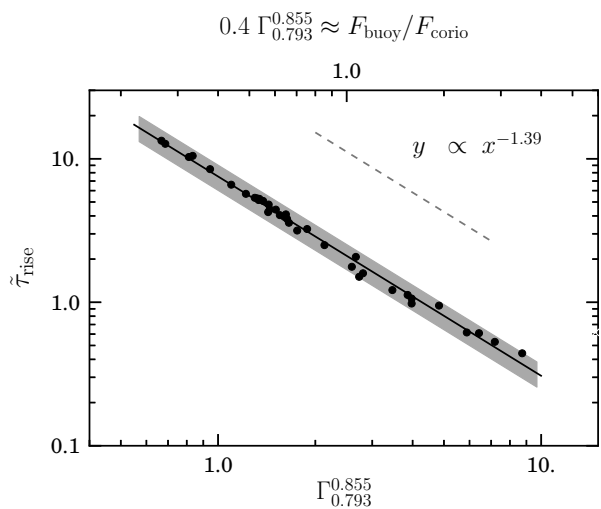}
  \caption{\label{fig:rrt-o-gamma-3D-dumfit} Relative {rise-time}  
           $\tilde{\tau}_{\rm rise}$ versus the scaling parameter  
           $\Gamma^{0.855}_{0.793}$. As in the 2D case{, } the function
           is self-similar, but with a different power $-1.39$.
          }
\end{figure}

{Fig.}~\ref{fig:rad-o-beta-mrot} shows the curvature radius versus
$\beta^{f_1} {\cal M}^{f_2}_{\rm rot}$, with $f_1$ and $f_2$ being
optimized to obtain the smallest residual of a linear fit.
%
%
  We find $f_1 = -0.161$ and $f_2 = -0.145$, which leads to $\alpha_1 = 0.793$
  and $\alpha_2 = 0.855$ (see Eq.~(\ref{eq:alphas-def}).
  As predicted in Sect.~\ref{sec:scaling-relation-of-rising-flux-tube}, $f_1$
  and $f_2$ are both negative, which shows that the curvature radius is indeed
  reduced for less buoyant flux tubes ($f_1 < 0$), and in faster
  rotating environments ($f_2 < 0$). It also gives us access to $\alpha_1$ and
  $\alpha_2$ which are required to test the validity of Eq.~(\ref{eq:rat-v-eq-rat-r}).
  To test this, we could verify that the relative rise-time, ${\tilde \tau}_{\rm rise}${, }
  scales with $\Gamma_{\alpha_1}^{\alpha_2}$.
  We computed from the simulations the values of $\alpha_1$ and
  $\alpha_2${, } and as illustrated in Fig.~\ref{fig:rrt-o-gamma-3D-dumfit}, the
  relative {rise-time} ${\tilde \tau}_{\rm rise}$ indeed scales with
  $\Gamma^{0.855}_{0.793}$.
  We can conclude that the assumption on which our prediction was
  based is appropriate.

This result has several implications: 
\begin{itemize}
  \item The radii of the fitted circles are good proxies for the
        curvature radii.  
  \item The magnetic tension indeed influences the regime of the rise. 
  \item It is possible to simulate the non-axisymmetric rise of
        compressible magnetic flux tubes for any solar-like star.
\end{itemize}
%

\subsection{Morphology of the flux tube and emergence}
\label{sec:5--studying-the-topology-of-the-flux-tube-and-emergence}

\begin{figure}[tbph]
  \centering
  \includegraphics[width=1.0\linewidth, trim=10 35 -15 15, clip]{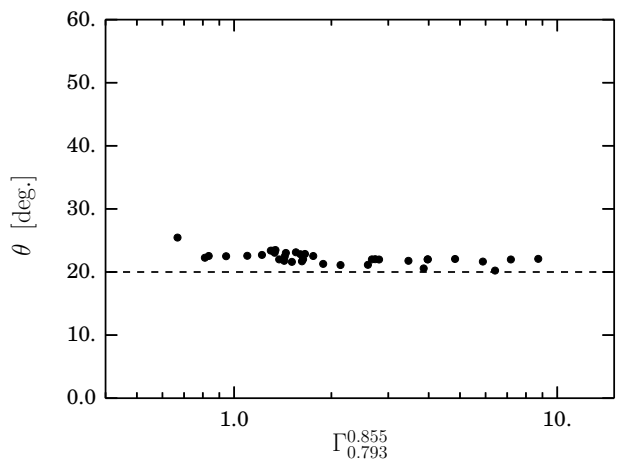}
  \caption{\label{fig:theta-o-gamma} Latitude of emergence versus the
           scaling parameter $\Gamma^{0.855}_{0.793}$. The dashed line
           represents the initial latitude for all simulations, namely
           $20^\circ$. 
          } 
\end{figure}

The morphology of the non-axisymmetric rise of
magnetic flux tubes has already been discussed for
anelastic simulations by \cite{Fan08} and \cite{Jouve+13}, and we show
here that our compressible simulations give similar results.  
In contrast to axisymmetric simulations, where magnetic flux tubes
rise radially or parallel to the rotation axis depending on the
regime of the rise (see Fig.~\ref{fig:paths}), non-axisymmetric flux
tubes take a more radial path independently of the regime (see
Figs.~\ref{fig:3D-rise} and \ref{fig:theta-o-gamma}).  
%
%

Alike the axisymmetric case, the angular momentum of the magnetic flux
tube has to be conserved. As the latter moves away from the rotation
axis, it decelerates. 
But according to \cite{Fan08}, because of its $\Omega$-shape the
non-axisymmetric rising magnetic flux tube builds up a 
pressure gradient between its apex and its feet. As a result{, } mass
flows along the tube's center line in the direction opposite to the
deceleration.
Non-axisymmetric flux tubes decelerate less than their axisymmetric
counterparts \citep[see][their Fig.~5]{Fan08} and the Coriolis
force reduces accordingly. As a result, non-axisymmetric flux tubes
always rise radially.  

\begin{figure}[tpbh]
  \centering
  \includegraphics[width=1.0\linewidth, trim=10 35 -15 15, clip]{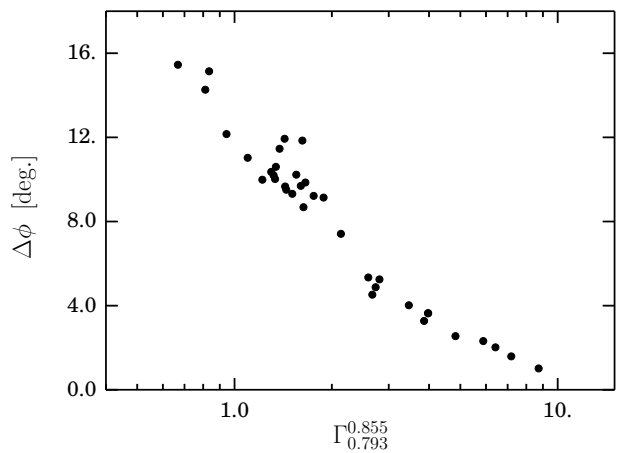}
  \caption{\label{fig:dphi-o-gamma} Azimuthal deflection angle between
           the initial azimuth of the buoyant part ($\phi_{\rm ini}$)
           and the azimuth of the emergence region against the scaling
           parameter $\Gamma^{0.855}_{0.793}$. 
           {We note} that the simulation labeled
           (f.\ref{fig:curv-radius}) in Table~\ref{tab:list-3D} is not
           plotted here because we lack emergence data.}  
\end{figure}

Furthermore, as shown by \cite{Jouve+13}, the asymmetry of the loop
increases with its azimuthal deflection{; } Fig.~\ref{fig:3D-rise} and 
Fig.~\ref{fig:dphi-o-gamma} show similar behavior.
%
%

These morphologic characteristics were already observed in former
anelastic studies. Therefore we conclude that our setup is reliable,
and we demonstrate that the anelastic approximation is able to
reproduce the morphology of compressible magnetic flux tubes.

\subsection{General relation for the {rise-time} in 3D}
\label{sec:5--extracting-the-general-relation-for-the-rising-flux-tube-in-3d}

In order to build a Babcock-Leighton dynamo that considers
non-instantaneous rises (a delayed Babcock-Leighton dynamo), we need a
relation between the relative {rise-time} ($\tilde{\tau}_{\rm{rise}}$)
and the initial strength of the magnetic flux tube. In
Fig.~\ref{fig:rrt-o-gamma-3D-dumfit} we plot the relative rise-time
over $\Gamma^{0.855}_{0.793}$. As in the axisymmetric setup the
relative {rise-time} follows a power law of $\Gamma^{\alpha_2}_{\alpha_1}$,
but in the non-axisymmetric case, with a different $\alpha_3$ of, {that is}: $-1.39$.   
%
\begin{equation}
  \label{eq:new-power-3D}
  \tilde{\tau}_{\rm{rise}} ~=~ 7.53 
  \left(\Gamma^{0.855}_{0.793}\right)^{-1.39} ~{\rm .}
\end{equation}
%
It is usually accepted that the relative {rise-time} is proportional to
the inverse square of the magnetic field strength, {but as we have just 
demonstrated, } this has to be reviewed. The STD-3D series shows that the
exponent depends on the azimuthal wavenumber $m$ of the initial
perturbation. For a given rotation rate ($\Omega$) we can write in
general
%
\begin{equation}
  \tilde{\tau}_{\rm{rise}} 
  \propto \left( \frac{B_{\phi}}{B_{\rm eq}} \right)^{\alpha_{\rm u}},
\end{equation}
%
where $\alpha_{\rm u} = \alpha_1 \alpha_2 \alpha_3$ is the universal exponent{, }
which depends on the azimuthal wavenumber $m$. The known values of
$\alpha_{\rm u}$ are summarized in Table~\ref{tab:alphau-values}. 
%
\begin{table}[tbh]
  \centering
  \caption{\label{tab:alphau-values} Values of the various exponents
           for both series, with an azimuthal wavenumber of $m=0$ and
           $m=8$.}  
  \begin{tabular}{ccccc}
    \hline\hline
                    & $\alpha_1$ & $\alpha_2$ & $\alpha_3$ & $\alpha_{\rm u}$\\
    \hline
    $m = 0$         &    $1.000$ &    $1.000$ &   $-2.00$ &   $-2.000$\\
    $m = 8$         &    $0.793$ &    $0.855$ &   $-1.39$ &   $-0.942$\\
    \hline
  \end{tabular}
\end{table}
%
In the particular case of $m=8$, the relative {rise-time} follows the
initial magnetic strength of the flux tube to the power $-0.942$. This
might have a significant impact on the solution of
delayed Babcock-Leighton dynamo.  

Again the plot of Fig.~\ref{fig:rrt-o-gamma-3D-dumfit} fails to
identify the nature of the regime. But as in the axisymmetric case we
can use a morphologic argument: panel ($c$) of Fig.~\ref{fig:3D-rise}
illustrates the less buoyant loop that remains symmetric until it 
reaches the surface. We suppose that in such a case 
both Coriolis {and} buoyant force are comparable. The regime becomes transitional for
$\Gamma^{0.855}_{0.793} \approx 2.5$ {and} therefore the proportionality
factor of Eq.~(\ref{eq:fratio-propto-gamma}) is close to $0.4$. We
indicated the resulting estimate of the force ratio on the upper axis
of Fig.~\ref{fig:rrt-o-gamma-3D-dumfit}. Any simulation with
$0.4\,\Gamma^{0.855}_{0.793} < 1$ is in a {rotation-dominated}
regime; simulations with a {larger-than-unity}
$0.4\,\Gamma^{0.855}_{0.793}$ are in a {buoyancy-dominated} regime.

Assuming that the morphological arguments we have used to
compute the proportionality factor are acceptable, we can identify the
regime of the rise. Fig.~\ref{fig:rrt-o-rom-2D-dumfit} and
\ref{fig:rrt-o-gamma-3D-dumfit} suggest that in the 
{rotation-dominated} domain non-axisymmetric flux {tubes} rise
faster than axisymmetric ones. But for a given 
{buoyancy-dominated} regime the axisymmetric flux tube will rise
faster.

These conclusions disagree with \cite{Fan08} where the
author concludes that in any case non-axisymmetric flux tubes rise
faster than axisymmetric ones. Thanks to the derivation in
Sect.~\ref{sec:scaling-relation-of-rising-flux-tube}, we can
demonstrate that the author compared flux tubes rising in different
regimes.
Indeed, we do agree with \cite{Fan08} that{, } for a given $\beta$ and a
given rotation rate ${\cal M}_{\rm rot}${, } a higher-$m$ tube will rise
faster, not because of non-axisymmetry, but because its regime has 
changed and tends towards a {buoyancy-dominated} regime. 

Additionally{, } we point out that the non-axisymmetric $\alpha_3$ is
closer to zero than the axisymmetric one. This aspect is beyond the
scope of the prediction made in
Sect.~\ref{sec:scaling-relation-of-rising-flux-tube} (the prediction
was made on $\alpha_1$ and $\alpha_2$ only). Nevertheless we would
like to suggest a physical interpretation. 
The pressure gradient that emerges between the apex and the feet of a
non-axisymmetric magnetic flux tube is independent of 
$\Gamma_{\alpha_1}^{\alpha_2}$, but depends on stratification and $m$. 
We suggest that $\alpha_3$ depends on the
competition between two {mechanisms}: the change in path taken by the
magnetic flux tube and the importance of the pressure gradient
building up between the apex and the feet of the flux tube. 
In the axisymmetric case the pressure gradient is absent, $\alpha_3$
is set exclusively by the variation of paths taken by the flux
tube. In the non-axisymmetric case there is no variation of the paths,
so the pressure gradient should be responsible for the value of
$\alpha_3$. This leads to the conclusion that in the
case of an $m=4$ mode the exponent should lie between $-2.0$ and
$-1.4$. This statement agrees with the conclusion drawn in
\cite{Moreno-insertis86}, where the author found an  
exponent of $-1.8$. To confirm this interpretation one should run
further simulations to extract the dependence of $\alpha_3$ on $m$.

\section{Discussion and conclusions}
\label{sec:discussions}

It is widely accepted that solar-like stars maintain their magnetic
field by a dynamo process. The choice of the dynamo process is
still heavily {debated}. In Babcock-Leighton (or flux-transport)
dynamos, the rise of magnetic flux tubes is an essential ingredient as
it links the strong toroidal fields of the tachocline with the 
near-surface source term of the dynamo.

In the various implementations of BL-dynamos, the {rise-time} of
magnetic flux tubes had been assumed to be instantaneous
until~\cite{Jouve+10}, who used a more complete description of the
{rise-time} of magnetic flux tubes and discussed the impact of the
resulting delayed generation of poloidal field on the dynamo. However,
the model describing the {rise-time} was rather simple. We {decided} to go 
further and {study} the {rise-time} of magnetic flux tubes in direct
numerical simulations. 

A scaling relation for the axisymmetric rise of thin flux tubes was
already proposed by~\cite{Schussler+92}. In the present work, we
carried out simulations of non-axisymmetrically rising magnetic flux
tubes in rotating, compressible, and stratified interiors. 
Non-axisymmetric perturbations to flux tubes lead to
the rise of $\Omega$-shaped loops. These loops have a much smaller
curvature radius than the corresponding torus in the axisymmetric
case. As a {result, } the tension force reduces notably the rise velocity
of the magnetic flux tube. 
We predicted that this difference changes the scaling relation of the
relative {rise-time} depending on the azimuthal wavenumber $m$ of the
initial perturbation. This wavenumber controls how strongly the
curvature radius is reduced. We present a theoretical approach
{describing} this phenomenon. From this theory{, } we predict that the
parameter controlling the regime of the rise, formerly introduced by
\cite{Choudhuri+87}, needs to be redefined as 
%
\begin{equation*}
  \Gamma^{\alpha_2}_{\alpha_1} = \left ( \frac{ v_{\rm A}^{\alpha_1} 
                                                c_{\rm s}^{1 - \alpha_1} }
                                              { \varpi \Omega } 
                                 \right)^{\alpha_2}.
\end{equation*}
%
This dimensionless parameter is proportional to the ratio of the
buoyant force over the Coriolis force acting on the flux tube. It
defines three regimes: {The} rotation dominated one when
$\Gamma^{\alpha_2}_{\alpha_1} < 1$, the transitional regime with
$\Gamma^{\alpha_2}_{\alpha_1} \approx 1${, } and the 
buoyancy dominated regime for 
$\Gamma^{\alpha_2}_{\alpha_1} > 1$. 

From two series of simulations we carried out with $m=0$ and $m=8$, we 
computed $\alpha_1$ and $\alpha_2$ in both cases and verified that the
regime of the rise is indeed controlled by
$\Gamma^{\alpha_2}_{\alpha_1}$. We then compared our compressible
simulations with the previous anelastic and {thin-flux-tube} studies. We
found that compressibility neither influences the morphology, nor the
{rise-time} of magnetic flux tubes.

Finally, we focused on our main goal, namely extracting the scaling
relation of the relative {rise-time} in light of the modified
scaling parameter. We found that in contrast to former
conclusions, for a given rotation rate $\Omega$, the relative rise
time does not necessarily scale with the inverse square of the
magnetic field, but that the power law depends on the azimuthal 
mode of the rising flux tube: 
%
\begin{equation*}
  {\tilde \tau}_{\rm rise} ~\propto~ \left( \Gamma_{\alpha_1}^{\alpha_2}
                                     \right)^{\alpha_3} ~~\rm{,}
\end{equation*}
%
which equals $(B_{\phi} / B_{\rm eq})^{\alpha_{\rm u}}$ for $\Omega = {\rm const}$.
For $m=0$ we confirm $\alpha_3 = -2$, while for $m=8$ we found
$\alpha_3 = -1.39$. This leads to {rise-times}that last less than 
$0.1~P_{\rm rot}$ for the {buoyancy-dominated} regime, about $P_{\rm rot}$ 
for the transitional regime, and more than $10~P_{\rm rot}$ 
for the {rotation-dominated} regime. This scaling relation can be applied to any main sequence
solar-like star, regardless of its rotation period and internal
magnetic fields. 

Nevertheless, the present model still neglects convection and
differential rotation. Our conclusions probably hold for flux tubes
being sufficiently magnetic to be unsensitive to convective
motion. But convection and differential rotation will influence the
above scaling relation. As shown by \cite{Weber+11}, weak flux tubes
will be advected by convective motions and will rise
faster. Nevertheless the present results are a step {toward} a general
description of flux rise that can possibly be tested using solar and
stellar cycle observations. 

Magnetic tension can explain the variation of $\alpha_1$
and $\alpha_2$, but does not address the variation of $\alpha_3$. 
Furthermore the fact that $\Gamma_{\alpha_1}^{\alpha_2}$ is a proxy for
the force ratio prevents this parameter from identifying the nature of
the regime in which the flux tube rises.
Explaining why the relative {rise-time} does not scale with 
the same power law of $\Gamma_{\alpha_1}^{\alpha_2}$ for various $m$,
and identifying the {dependencies} of the proportionality factor on
viscosity, twist, and thermal conduction, represent a numerical
challenge and will require further simulations. 

With a complete theory we will be able to construct a universal
model for the {rise-time} of magnetic flux tubes in low-mass stars, but
the present conclusions are already sufficient to enrich the idea of
delayed Babcock-Leighton dynamos.

\begin{acknowledgements}
We would like to thank Laur\`ene Jouve, J\"orn Warnecke, Sydney Barnes
and Federico Spada for constructive discussions and comments. This
work was supported by the DFG grant Ar 355/9-1.
\end{acknowledgements}

\bibliographystyle{aa}
\bibliography{references}

\begin{thebibliography}{59}
\expandafter\ifx\csname natexlab\endcsname\relax\def\natexlab#1{#1}\fi

\bibitem[{Babcock(1962)}]{Babcock62}
Babcock, H.~W. 1962, Transactions of the International Astronomical Union,
  Series B, 11, 419

\bibitem[{Brown {et~al.}(2010)Brown, Browning, Brun, Miesch, \&
  Toomre}]{Brown+10}
Brown, B.~P., Browning, M.~K., Brun, A.~S., Miesch, M.~S., \& Toomre, J. 2010,
  \apj, 711, 424

\bibitem[{Browning \& Priest(1983)}]{Browning+83}
Browning, P.~K. \& Priest, E.~R. 1983, \apj, 266, 848

\bibitem[{Brun {et~al.}(2004)Brun, Miesch, \& Toomre}]{Brun+04}
Brun, A.~S., Miesch, M.~S., \& Toomre, J. 2004, \apj, 614, 1073

\bibitem[{Caligari {et~al.}(1995)Caligari, Moreno-Insertis, \&
  Sch\"ussler}]{Caligari+95}
Caligari, P., Moreno-Insertis, F., \& Sch\"ussler, M., M. 1995, \apj, 441, 886

\bibitem[{Caligari {et~al.}(1996)Caligari, Sch\"ussler, Solanki, Schaerer, \&
  Stix}]{Caligari+96}
Caligari, P., Sch\"ussler, M., Solanki, S.~K., Schaerer, D., \& Stix, M. 1996,
  \aplettcomm, 34, 17

\bibitem[{Caligari {et~al.}(1994)Caligari, Sch\"ussler, Stix, \&
  Solanki}]{Caligari+94}
Caligari, P., Sch\"ussler, M., Stix, M., \& Solanki, S.~K. 1994, in Cool
  {Stars}, {Stellar} {Systems}, and the {Sun}, ed. J.-P. Caillault, Vol.~64,
  387

\bibitem[{Cattaneo \& Hughes(1987)}]{Cattaneo+87}
Cattaneo, F. \& Hughes, D.~W. 1987, in {NASA} {Conference} {Publication}, ed.
  G.~Athay \& D.~S. Spicer, Vol. 2483, 101--104

\bibitem[{Cattaneo {et~al.}(1989)Cattaneo, Tzihong, \& Hughes}]{Cattaneo+89}
Cattaneo, F., Tzihong, C., \& Hughes, D.~W. 1989, \baps, 34, 1294

\bibitem[{Charbonneau(2010)}]{Charbonneau10}
Charbonneau, P. 2010, \lrsp, 7, 3

\bibitem[{Charbonneau(2014)}]{Charbonneau14}
Charbonneau, P. 2014, Annual Review of Astronomy and Astrophysics, 52, 251

\bibitem[{Cheung \& Isobe(2014)}]{Cheung+14}
Cheung, M. C.~M. \& Isobe, H. 2014, Living Reviews in Solar Physics, 11

\bibitem[{Cheung {et~al.}(2006)Cheung, Moreno-Insertis, \&
  Sch\"ussler}]{Cheung+06}
Cheung, M. C.~M., Moreno-Insertis, F., \& Sch\"ussler, M. 2006, \aap, 451, 303

\bibitem[{Chou \& Fisher(1989)}]{Chou+89}
Chou, D.-Y. \& Fisher, G.~H. 1989, \apj, 341, 533

\bibitem[{Choudhuri \& D'Silva(1990)}]{Choudhuri+90}
Choudhuri, A.~R. \& D'Silva, S. 1990, \aap, 239, 326

\bibitem[{Choudhuri \& Gilman(1987)}]{Choudhuri+87}
Choudhuri, A.~R. \& Gilman, P.~A. 1987, \apj, 316, 788

\bibitem[{DeLuca {et~al.}(1997)DeLuca, Fan, \& Saar}]{Deluca+97}
DeLuca, E.~E., Fan, Y., \& Saar, S.~H. 1997, \apj, 481, 369

\bibitem[{Fan(2001)}]{Fan01}
Fan, Y. 2001, \apj, 546, 509

\bibitem[{Fan(2008)}]{Fan08}
Fan, Y. 2008, \apj, 676, 680

\bibitem[{Fan {et~al.}(2013)Fan, Featherstone, \& Fang}]{Fan+13}
Fan, Y., Featherstone, N., \& Fang, F. 2013, ArXiv e-prints

\bibitem[{Fan {et~al.}(1994)Fan, Fisher, \& McClymont}]{Fan+94}
Fan, Y., Fisher, G.~H., \& McClymont, A.~N. 1994, \apj, 436, 907

\bibitem[{Fan {et~al.}(1998)Fan, Zweibel, \& Lantz}]{Fan+98}
Fan, Y., Zweibel, E.~G., \& Lantz, S.~R. 1998, \apj, 493, 480

\bibitem[{Granzer(2004)}]{Granzer04}
Granzer, T. 2004, Astronomische Nachrichten, 325, 417

\bibitem[{Granzer {et~al.}(2000)Granzer, Sch\"ussler, Caligari, \&
  Strassmeier}]{Granzer+00}
Granzer, T., Sch\"ussler, M., Caligari, P., \& Strassmeier, K.~G. 2000, \aap,
  355, 1087

\bibitem[{Hughes {et~al.}(1997)Hughes, Wissink, Matthews, \&
  Proctor}]{Hughes+97}
Hughes, D.~W., Wissink, J.~G., Matthews, P.~C., \& Proctor, M. R.~E. 1997, in
  1st {Advances} in {Solar} {Physics} {Euroconference}. {Advances} in {Physics}
  of {Sunspots}, ed. B.~Schmieder, J.~C. del Toro~Iniesta, \& M.~Vazquez, Vol.
  118, 66

\bibitem[{Jouve \& Brun(2009)}]{Jouve+09}
Jouve, L. \& Brun, A.~S. 2009, \apj, 701, 1300

\bibitem[{Jouve {et~al.}(2013)Jouve, Brun, \& Aulanier}]{Jouve+13}
Jouve, L., Brun, A.~S., \& Aulanier, G. 2013, \apj, 762, 4

\bibitem[{Jouve {et~al.}(2010)Jouve, Proctor, \& Lesur}]{Jouve+10}
Jouve, L., Proctor, M. R.~E., \& Lesur, G. 2010, \aap, 519, A68

\bibitem[{K\"apyl\"a {et~al.}(2010)K\"apyl\"a, Korpi, Brandenburg, Mitra, \&
  Tavakol}]{Kapyla+10}
K\"apyl\"a, P.~J., Korpi, M.~J., Brandenburg, A., Mitra, D., \& Tavakol, R.
  2010, \an, 331, 73

\bibitem[{K\"apyl\"a {et~al.}(2012)K\"apyl\"a, Mantere, \&
  Brandenburg}]{Kapyla+12}
K\"apyl\"a, P.~J., Mantere, M.~J., \& Brandenburg, A. 2012, \apjl, 755, L22

\bibitem[{Leighton(1969)}]{Leighton69}
Leighton, R.~B. 1969, The Astrophysical Journal, 156, 1

\bibitem[{Longcope \& Klapper(1997)}]{Longcope+97}
Longcope, D.~W. \& Klapper, I. 1997, \apj, 488, 443

\bibitem[{Matthews {et~al.}(1995)Matthews, Hughes, \& Proctor}]{Matthews+95}
Matthews, P.~C., Hughes, D.~W., \& Proctor, M. R.~E. 1995, \apj, 448, 938

\bibitem[{Miesch(2005)}]{Miesch05}
Miesch, M.~S. 2005, \lrsp, 2, 1

\bibitem[{Moreno-Insertis(1983)}]{Moreno-insertis83}
Moreno-Insertis, F. 1983, \aap, 122, 241

\bibitem[{Moreno-Insertis(1986)}]{Moreno-insertis86}
Moreno-Insertis, F. 1986, \aap, 166, 291

\bibitem[{Moreno-Insertis(1997)}]{Moreno-insertis+97}
Moreno-Insertis, F. 1997, in 1st {Advances} in {Solar} {Physics}
  {Euroconference}. {Advances} in {Physics} of {Sunspots}, ed. B.~Schmieder,
  J.~C. del Toro~Iniesta, \& M.~Vazquez, Vol. 118, 45--65

\bibitem[{Moreno-Insertis \& Emonet(1996)}]{Moreno-insertis+96}
Moreno-Insertis, F. \& Emonet, T. 1996, \apjl, 472, L53

\bibitem[{Moreno-Insertis {et~al.}(1992)Moreno-Insertis, Sch\"ussler, \&
  Ferriz-Mas}]{Moreno-insertis+92}
Moreno-Insertis, F., Sch\"ussler, M., \& Ferriz-Mas, A. 1992, \aap, 264, 686

\bibitem[{Nelson {et~al.}(2014)Nelson, Brown, Sacha~Brun, Miesch, \&
  Toomre}]{Nelson+14}
Nelson, N.~J., Brown, B.~P., Sacha~Brun, A., Miesch, M.~S., \& Toomre, J. 2014,
  \solphys, 289, 441

\bibitem[{Parker(1955)}]{Parker55}
Parker, E.~N. 1955, \apj, 122, 293

\bibitem[{Pinto \& Brun(2013)}]{Pinto+13}
Pinto, R.~F. \& Brun, A.~S. 2013, \apj, 772, 55

\bibitem[{Racine {et~al.}(2011)Racine, Charbonneau, Ghizaru, Bouchat, \&
  Smolarkiewicz}]{Racine+11}
Racine, E., Charbonneau, P., Ghizaru, M., Bouchat, A., \& Smolarkiewicz, P.~K.
  2011, \apj, 735, 46

\bibitem[{R\"adler {et~al.}(2003)R\"adler, Kleeorin, \&
  Rogachevskii}]{Radler+03}
R\"adler, K.-H., Kleeorin, N., \& Rogachevskii, I. 2003, \gapfd, 97, 249

\bibitem[{Rempel \& Cheung(2014)}]{Rempel+14}
Rempel, M. \& Cheung, M. C.~M. 2014, \apj, 785, 90

\bibitem[{Sch\"ussler(1980)}]{Schussler80}
Sch\"ussler, M. 1980, Nature, 288, 150

\bibitem[{Sch\"ussler \& Rempel(2002)}]{Schussler+02}
Sch\"ussler, M. \& Rempel, M. 2002, in From {Solar} {Min} to {Max}: {Half} a
  {Solar} {Cycle} with {SOHO}, ed. A.~Wilson, Vol. 508, 499--506

\bibitem[{Sch\"ussler \& Solanki(1992)}]{Schussler+92}
Sch\"ussler, M. \& Solanki, S.~K. 1992, \aap, 264, L13

\bibitem[{Spruit(1981)}]{Spruit81}
Spruit, H.~C. 1981, \aap, 102, 129

\bibitem[{Spruit \& van Ballegooijen(1982)}]{Spruit+82}
Spruit, H.~C. \& van Ballegooijen, A.~A. 1982, \aap, 106, 58

\bibitem[{Strassmeier(2009)}]{Strassmeier09}
Strassmeier, K.~G. 2009, Astronomy and Astrophysics Review, 17, 251

\bibitem[{van Ballegooijen(1983)}]{Van_ballegooijen83}
van Ballegooijen, A.~A. 1983, \aap, 118, 275

\bibitem[{Warnecke {et~al.}(2013)Warnecke, Losada, Brandenburg, Kleeorin, \&
  Rogachevskii}]{Warnecke+13}
Warnecke, J., Losada, I.~R., Brandenburg, A., Kleeorin, N., \& Rogachevskii, I.
  2013, \apjl, 777, L37

\bibitem[{Weber \& Fan(2015)}]{Weber+15}
Weber, M.~A. \& Fan, Y. 2015, ArXiv e-prints

\bibitem[{Weber {et~al.}(2011)Weber, Fan, \& Miesch}]{Weber+11}
Weber, M.~A., Fan, Y., \& Miesch, M.~S. 2011, \apj, 741, 11

\bibitem[{Wissink {et~al.}(2000)Wissink, Hughes, Matthews, \&
  Proctor}]{Wissink+00}
Wissink, J.~G., Hughes, D.~W., Matthews, P.~C., \& Proctor, M. R.~E. 2000,
  \mnras, 318, 501

\bibitem[{Yoshimura(1985)}]{Yoshimura85}
Yoshimura, H. 1985, \pasj, 37, 171

\bibitem[{Ziegler(2011)}]{Ziegler11}
Ziegler, U. 2011, \jcphys, 230, 1035

\bibitem[{Ziegler(2012)}]{Ziegler12}
Ziegler, U. 2012, SIAM J. Sci. Comput., 34, C102

\end{thebibliography}

\begin{appendix}

\section{Adiabatically stratified atmosphere}
\label{sec-8-1}

In this section, we derive the profiles of the three
hydrodynamical quantities $T$, $\rho${, } and $P$ for an adiabatically 
stratified atmosphere. These profiled are used to mimic a stellar
convective zone avoiding convective motions but still reproducing a
realistic stellar stratification. In such a case we have to presume
that the luminosity of the star is transported exclusively by
radiation: 
%
\[ \vec{F}_{\rm rad} = - \kappa \frac{dT}{dr} \vec{e}_r\,,\]
%
Here,
%
\[ |\vec{F}_{\rm rad}| = \frac{L}{4\pi r^2}.\]
%
Hence,
\[ \frac{dT}{dr} = - \frac{L}{4\pi r^2} \frac{1}{\kappa}. \]
%
Integrating over radius we obtain a function for $T(r)$,
%
\[ T(r) = \frac{L}{4\pi \kappa} \frac{1}{r} + const.\]
%
Setting the following boundary condition: $T(R_0) = T_0$, 
%
\[ T(r) = T_0 + \frac{L}{4\pi \kappa} \left(\frac{1}{r} 
              - \frac{1}{R_0}\right) ~{\rm .}\]
%
In that case the luminosity and $\kappa$ are the input parameters, but
we need to control the stratification. Hence, we rewrite the two
latter variables as functions of more meaningful quantities, namely
$\nabla$ (the logarithmic temperature gradient) and {pressure-scale}
height. 
Moreover the {pressure-scale}
height at the surface is a useful length for our problem since: 
%
\[       H_{\!P_0} = P_0 
                   ~ \left( 
                           \left. \frac{dP}{dr} \right|_{R_0} 
                     \right)^{-1} ~{\rm .} \]
%
Moreover the stellar interior is in hydrostatic equilibrium, {where}
%
\[ \frac{dP}{dr} = - \rho \frac{GM}{r^2} ~{\rm ,}\]
%
which leads to a convenient definition of the {pressure-scale} height:
%
\[ H_{\!P_{0}} = - \frac{P_{0}}{\rho_0} \frac{R_0^2}{GM} ~{\rm .}\] 
%
With both $\nabla$ and $H_{\!P_{0}}$, we rewrite $T(r)$ in
a more convenient way. We first relate $\nabla$ and $H_{\!P_{0}}$.
%
\[ \nabla = \frac{d \ln T}{d \ln P}   \]
\[ \nabla = \frac{dT}{T} \frac{P}{dP} \]
\[ \nabla = \frac{P}{T} \frac{dT}{dr} \frac{dr}{dP} ~{\rm ,}\]
%
{which gives},
%
\[ \nabla = - \frac{P}{T\rho} \frac{dT}{dr} \frac{r^2}{GM} ~{\rm ,}\]
%
and considering the ideal gas equation of state:
%
\[ P = \frac{\rho kT}{m\mu} ~{\rm ,}\]
\[ \frac{P}{T\rho} = \frac{k}{m\mu} = Cst = \frac{P_0}{T_0 \rho_0} ~{\rm .}\]
%
Finally,
%
\[ \nabla = - \frac{P_0}{T_0 \rho_0} \frac{dT}{dr} \frac{r^2}{GM}, \]
%
where we immediately identify $H_{\!P_{0}}$ and rewrite
$\frac{dT}{dr}$ as a function of both:
%
\[ \frac{dT}{dr} = - T_{0} \frac{\nabla}{H_{\!P_{0}}} 
                     \left( \frac{R_0}{r} \right)^2  ~{\rm .}\]
%
By integration we obtain a convenient temperature profile, where we are
able to control both the {pressure-scale} height and the adiabaticity of
our layer, 
%
\[ T(r) = T_0 \left[ 1 + \frac{\nabla R_0^2}{H_{\!P_{0}}} 
              \left(\frac{1}{r} - \frac{1}{R_0}\right) \right] ~{\rm .}\]
%
Since $\nabla$ is a constant all over the domain it also has the
advantage of relating $T$ and $P$: 
%
\[ \nabla = \frac{d\ln T}{d\ln P} = \frac{\ln(T/T_0)}{\ln(P/P_0)} ~{\rm, }\]
%
{leading to}
%
\[ P(r) = P_0 \left(\frac{T(r)}{T_0}\right)^{1/\nabla}. \]
%
Reusing the equation of state it becomes clear that:
%
\[ \rho(r) = \rho_0 \left(\frac{T(r)}{T_0}\right)^{1/\nabla-1} ~{\rm .}\]

\clearpage
\onecolumn

\section{List of simulations}
\label{sec-8-2}

\begin{table}[tbh]
  \caption{\label{tab:list-of-sim2D}List of all axisymmetric
           simulations based on the STD-2D setup.}
  \centering
  \begin{tabular}{ccr@{.}lcc}
    \multicolumn{2}{l}{\textbf{Series 2D}} & \multicolumn{2}{l}{$ m = 0 $} & $ \alpha_1 = 1 $ & $ \alpha_2 = 1 $ \\[1.ex]
    \hline
    \hline\\[-8pt]
    $\Gamma^{\alpha_2}_{\alpha_1}$ & $\cal{M}_{\rm{rot}}$ & \multicolumn{2}{c}{$\beta$} & $\tilde{\tau}_{\rm{rise}}$ & labels \\[.5ex]
    \hline\\[-8pt]
    3.162 & 0.325 &  1&135 & 0.254 &\\
    2.846 & 0.325 &  1&401 & 0.260 &\\
    2.846 & 0.244 &  2&491 & 0.261 &\\
    2.372 & 0.406 &  1&291 & 0.359 &\\
    2.372 & 0.325 &  2&018 & 0.393 &\\
    1.897 & 0.488 &  1&401 & 0.477 &\\
    1.897 & 0.325 &  3&153 & 0.550 & ($b$) \\
    1.708 & 0.488 &  1&730 & 0.651 &\\
    1.660 & 0.325 &  4&118 & 0.814 &\\
    1.550 & 0.406 &  3&026 & 1.082 &\\
    1.470 & 0.406 &  3&360 & 1.325 &\\
    1.455 & 0.488 &  2&384 & 1.276 & ($c$) \\
    1.360 & 0.406 &  3&929 & 1.488 &\\
    1.297 & 0.488 &  3&001 & 1.676 &\\
    1.265 & 0.406 &  4&540 & 1.732 &\\
    1.217 & 0.650 &  1&914 & 1.772 &\\
    1.217 & 0.569 &  2&501 & 1.713 &\\
    1.217 & 0.528 &  2&900 & 1.811 &\\
    1.217 & 0.488 &  3&404 & 1.778 &\\
    1.217 & 0.447 &  4&050 & 1.813 & ($a.1$)\\
    1.217 & 0.406 &  4&901 & 1.813 & ($a.2$)\\
    1.217 & 0.366 &  6&051 & 1.779 &\\
    1.217 & 0.325 &  7&658 & 1.775 &\\
    1.217 & 0.285 & 10&00  & 1.723 &\\
    1.217 & 0.203 & 19&60  & 1.670 &\\
    1.217 & 0.163 & 30&63  & 1.653 &\\
    1.217 & 0.122 & 54&45  & 1.621 &\\
    1.186 & 0.488 &  3&587 & 1.876 &\\
    1.186 & 0.406 &  5&166 & 1.815 &\\
    1.186 & 0.325 &  8&072 & 1.777 &\\
    1.154 & 0.406 &  5&453 & 1.895 &\\
    1.107 & 0.488 &  4&118 & 1.979 &\\
    1.059 & 0.569 &  3&303 & 2.072 &\\
    0.791 & 0.325 & 18&16  & 4.460 & ($d$) \\
    \hline
  \end{tabular}
  \tablefoot{The labels refer to the panels in Fig. \ref{fig:paths},
    with ($a.1$) and ($a.2$) referring to the green and red contours of
    panel ($a$), respectively.}
\end{table}

\begin{table}[tbh]
  \caption{\label{tab:list-3D}List of all non-axisymmetric simulations
           based on the STD-3D setup.}
  \centering
  \begin{tabular}{r@{.}lcr@{.}lr@{.}lc}
    \multicolumn{2}{l}{\textbf{Series 3D}} & \multicolumn{2}{l}{$m = 8$} & \multicolumn{3}{l}{$\alpha_1 = 0.793$} &  \multicolumn{1}{l}{$\alpha_2 = 0.855$} \\[1ex]
    \hline
    \hline
    \multicolumn{3}{l}{} & \multicolumn{2}{l}{} & \multicolumn{2}{l}{} & \\[-8pt]
    \multicolumn{2}{c}{$\Gamma^{\alpha_2}_{\alpha_1}$} & $\cal{M}_{\rm{rot}}$ & \multicolumn{2}{c}{$\beta$} & \multicolumn{2}{c}{$\tilde{\tau}_{\rm{rise}}$} & labels \\[.5ex]
    \hline
    \multicolumn{3}{l}{} & \multicolumn{2}{l}{} & \multicolumn{2}{l}{} & \\[-8pt]
    8&733 &  0.08 &  1&166 &  0&441 & ($a.11$) \\
    7&185 &  0.08 &  2&073 &  0&53  & \\
    6&419 &  0.04 & 16&605 &  0&609 & \\
    5&885 &  0.08 &  3&735 &  0&616 & \\
    4&828 &  0.16 &  1&166 &  0&948 & ($a.12$)/($b.11$)\\
    3&973 &  0.16 &  2&073 &  0&98  & \\
    3&973 &  0.16 &  2&073 &  1&062 & \\
    3&86  &  0.08 & 12&957 &  1&125 & \\
    3&463 &  0.16 &  3&108 &  1&216 & \\
    2&809 & 0.241 &  2&073 &  1&593 & ($b.12$)\\
    2&733 & 0.281 &  1&523 &  1&505 & \\
    2&669 & 0.321 &  1&166 &  2&072 & \\
    2&594 &  0.16 &  7&288 &  1&768 & ($c.11$) \\
    2&134 &  0.16 & 12&957 &  2&502 & \\
    1&886 &  0.16 & 18&658 &  3&24  & \\
    1&757 & 0.481 &  1&44  &  3&162 & ($c.12$)\\
    1&657 & 0.481 &  1&713 &  3&582 & \\
    1&634 & 0.241 & 10&237 &  3&816 & \\
    1&621 &  0.16 & 29&153 &  4&1   & \\
    1&604 & 0.401 &  2&985 &  3&94  & \\
    1&553 & 0.481 &  2&073 &  4&05  & \\
    1&509 & 0.241 & 12&957 &  4&425 & \\
    1&446 & 0.481 &  2&559 &  4&44  & \\
    1&434 & 0.321 &  7&288 &  4&8   & \\
    \textit{1}&\textit{429} & \textit{0.196} & \textit{25}&\textit{374} &  \textit{4}&\textit{251} & \\
    1&378 & 0.241 & 16&923 &  5&085 & \\
    1&343 & 0.561 &  2&159 &  5&257 & \\
    1&335 & 0.481 &  3&239 &  5&16  & \\
    1&321 & 0.561 &  2&265 &  5&299 & \\
    1&299 & 0.561 &  2&38  &  5&348 & ($d.12$)\\
    1&219 & 0.481 &  4&231 &  5&7   & \\
    1&098 & 0.481 &  5&759 &  6&6   & ($d.11$) \\ 
    0&944 & 0.561 &  6&092 &  8&47  & \\
    \textit{0}&\textit{834} & \textit{0.481} & \textit{12}&\textit{957} & \textit{10}&\textit{482} & \\
    \textit{0}&\textit{812} & \textit{0.561} &  \textit{9}&\textit{519} & \textit{10}&\textit{29}  & ($e.12$) \\ 
    \textit{0}&\textit{686} & \textit{0.481} & \textit{23}&\textit{034} & \textit{12}&\textit{72}  & ($f.12$) \\ 
    \textit{0}&\textit{668} & \textit{0.561} & \textit{16}&\textit{923} & \textit{13}&\textit{37}  & \\
    \hline
  \end{tabular}
  \tablefoot{The letters of the labels refer to the panels in
             Fig.~\ref{fig:3D-rise} and
             Fig.~\ref{fig:curv-radius}, respectively referred by their 
             number. The lines in italics refer to the simulations 
             where the computation of the curvature radius is not reliable.}
\end{table}

\end{appendix}

\end{document}